\newif\ifprintfig
\printfigtrue

\documentclass[numberedappendix]{emulateapj}
\usepackage{graphicx}

\newcommand\pc{{\rm\,pc}}
\newcommand\cm{{\rm\,cm}}
\newcommand\kpc{{\rm\,kpc}}

\newcommand\Myr{{\rm\,Myr}}
\newcommand\Gyr{{\rm\,Gyr}}
\newcommand\yrs{{\rm\,yrs}}
\newcommand\kmsec{{\rm\,km\,s^{-1}}}
\newcommand\kms{\kmsec}

\newcommand\s{{\rm\,s}}
\newcommand\K{{\rm\,K}}

\newcommand\Msun{{\rm\,M_\odot}}
\newcommand\msun{{\rm\,M_\odot}}

\newcommand\clock{\count0=\time \divide\count0 by 60
     \count1=\count0 \multiply\count1 by -60 \advance\count1 by \time
     \number\count0:\ifnum\count1<10{0\number\count1}\else\number\count1\fi}

\shortauthors{Dalcanton}
\shorttitle{Gas Flows in Galaxies}

\begin{document}

\title{The Metallicity of Galaxy Disks: Infall versus Outflow}

\author{Julianne J. Dalcanton\altaffilmark{1,2}}
\affil{Department of Astronomy, University of Washington, Box 351580,
Seattle WA, 98195}

\altaffiltext{1}{e-mail address: jd@astro.washington.edu}
\altaffiltext{2}{Alfred P.\ Sloan Foundation Fellow}
  
\begin{abstract}

  Galaxies are not ``closed-boxes''.  Both gas accretion (infall) and
  winds (outflow) change a galaxy's metallicity and gas fraction,
  lowering the galaxy's effective yield
  ($\equiv\!Z/\ln{(f_{gas}^{-1})}$).  Low effective yields are seen in
  galaxies with rotation speeds less than $\sim\!120\kms$, and have
  been widely interpreted as the onset of supernova-driven winds below
  a characteristic galaxy mass.  However, accretion of metal-poor gas
  is also a viable explanation for the low effective yields observed
  in low mass galaxies.  Analytic calculations presented in this paper
  prove: (1) that neither infall nor unenriched outflows can produce
  effective yields that are as low as observed; (2) that
  metal-enriched outflows are the only mechanism that can
  significantly reduce the effective yield, but only in gas-rich
  systems; (3) that it is nearly impossible to reduce the effective
  yield of a gas-poor system, no matter how much gas is lost or
  accreted; and (4) that any subsequent star formation rapidly drives
  the effective yield back to the closed-box value.  Given that
  massive galaxy disks are systematically more gas-poor than low mass
  dwarf irregulars, these calculations imply that: (1) massive
  galaxies may have experienced substantial infall and/or outflow in
  spite of their high effective yields; (2) only gas-rich systems can
  have low effective yields; and (3) only systems with low star
  formation rates can maintain low effective yields in between
  episodes of mass loss.  The drop in effective yield seen in low mass
  galaxies is therefore not necessarily due to the onset of supernova-driven
  winds.  Instead, it is likely due to the fact that galaxies with
  $V_c\lesssim\!120\kms$ have surface densities lying entirely below
  the Kennicutt star formation threshold, as previously noted by Verde
  et al. and Dalcanton et al.  The resulting low star formation
  efficiency keeps these galaxies sufficiently gas-rich that their
  effective yields can be reduced by outflows, and keeps
  their effective yields low between episodes of gas loss.  Additional
  calculations confirm that there is no galaxy mass at which the
  metal-loss becomes dramatically stronger.  The fraction of baryonic
  mass lost through winds varies only weakly with galaxy mass, shows
  no sharp upturn at any mass scale, and does not require that more
  than 15\% of baryons have been lost by galaxies of any mass.
  Supernova feedback is therefore unlikely to be an effective
  mechanism for removing large amounts of gas from low mass disk
  galaxies.  In addition, the dependence between metal-loss and galaxy
  mass is sufficiently weak that massive galaxies dominate metal
  enrichment of the IGM.  The calculations in this paper are based on
  impulsive accretion, outflow, and star formation events that provide
  limiting cases for more arbitrary chemical evolution histories, as
  proven in an Appendix.

\end{abstract}
\keywords{galaxies: formation --- ISM: evolution --- galaxies: abundances
galaxies: ISM --- galaxies: evolution}

\section{Introduction}  \label{introsec}

Since the seminal work of \citet{larson74}, supernova-driven winds
have been a favored mechanism for explaining the properties of low
mass galaxies.  Modern theories of galaxy formation frequently invoke
gas outflows (or ``feedback'') to explain the lack of gas in dwarf
spheroidals \citep[e.g.,][]{sandage65,dekel86,lanfranchi04}, the
paucity of low mass galaxies compared to CDM simulations
\citep[e.g.,][]{white91}, the hot gaseous halos around dwarf
starbursts \citep[e.g.,][]{marlowe95, martin98, martin02, ott05}, and the
metal enrichment of the intergalactic medium
\citep[e.g.,][]{silk87,madau01,mori02}.

Outflows may also be partially responsible for low metallicities in
disk galaxies, particularly in low mass dwarf irregulars.  For many
years there has been evidence for a mass-metallicity relationship
among galaxies, with low mass galaxies having systematically lower
metallicities \citep[see Figure~7 of the review by ][ for an early
plot of this relation\footnote{While outflow was initially considered
  as a possible explanation for the mass-metallicity relationship in
  disks, most initial theoretical work on outflows focused primarily
  on elliptical systems \citep[e.g.][]{hartwick80}.}]{pagel81}.  While
some of the correlation between mass and metallicity is certainly due
to less past enrichment in gas-rich low mass galaxies, some may also
be due to a direct loss of metals through winds.

Recently, after bringing abundance measurements to a common
metallicity scale and fiducial radius, \citet{garnett02} found a tight
relationship between metallicity and galaxy mass.  By combining
measurements of the metallicity with estimates of the gas richness, he
showed that galaxies systematically depart from the closed-box model
of chemical evolution below a characteristic mass scale near
$V_c\sim125\kms$.  These results have since been qualitatively
confirmed by \citet{pilyugin04}, using an alternative abundance
calibration, and by \citet{tremonti04}, using a much larger sample of
metallicities from the Sloan Digital Sky Survey (SDSS), although with
less accurate gas mass fractions.  These results have widely been
interpreted as evidence for the onset of strong galactic winds at a
specific galaxy mass scale, below which gas and metals can easily
escape the shallow potential well.

Although most theoretical attention has focused on outflow, infall is
an equally appealing mechanism for producing low metallicities.
Outflow reduces the metallicity by preferentially driving metals out
of a system, while infall reduces the metallicity by diluting a
galaxy's interstellar medium (ISM) with fresh, low metallicity gas.
As shown by \citet{koppen99}, the metallicity drops if rate of infall
is higher than the rate of star formation, thus allowing the accreted
metal-poor gas to dilute the ISM faster than it can be enriched by
evolving stars.  Given the very low star formation rates and star
formation efficiencies seen in the majority of low mass disk galaxies
\citep[e.g.][]{hunter04}, infall is therefore a viable mechanism for
producing low metallicities in low mass galaxies.

The conclusion that low mass galaxies have not evolved as
``closed-boxes'' is based on measurements of their effective yields.
The effective yield measures how a galaxy's metallicity deviates from
what would be expected for a galaxy that had the same gas mass
fraction, but that had been evolving as a closed-box, i.e.\ with no
inflow or outflow of gas.  A system that evolves as a closed-box
obeys a simple analytic relationship between the metallicity of the
gas and the gas mass fraction (see reviews by Pagel 1997 and Tinsley
1980).  As gas is converted into stars, the gas mass fraction
$f_{gas}$ ($\equiv M_{gas}/(M_{gas}+M_{stars})$) decreases and the
metallicity $Z_{gas}$ of the gas increases according to
\begin{equation}    \label{closedboxeqn}
Z_{gas} = y_{true} \, \ln{(1/f_{gas})}
\end{equation}
\citep{searle72} where $y_{true}$ is the true nucleosynthetic yield,
defined as the mass in primary elements freshly produced by massive
stars, in units of the stellar mass that is locked-up in long-lived
stars and stellar remnants.  For closed-box evolution, the
metallicity increases without limit, and equals the nucleosynthetic
yield when $f_{gas}=\exp{(-1)}$.  Equation~\ref{closedboxeqn} assumes
that the metals produced by a generation of stars are instantly
returned to the ISM and are well-mixed with existing gas.  This
``instantaneous recycling'' approximation is likely to be valid for
measurements of galaxies' current gas-phase metallicities, which
typically use oxygen abundances measured in HII regions.  Since the
production of oxygen is dominated by winds and SN from massive stars
($>\!8\msun$; see Figures~5~\&~7 from Maeder 1992), prompt return of
oxygen to the local ISM is a reasonable assumption.

If a galaxy evolves as a closed-box, the ratio of
$Z_{gas}/\ln{f_{gas}^{-1}}$ should be a constant equal to the
nucleosynthetic yield.  However, this ratio will be lower if metals
have been lost from the system through winds, or if the current gas
has been diluted with fresh infall of metal-poor gas.  One may
therefore define the above ratio as the ``effective yield'' $y_{eff}$
\begin{equation}    \label{peffeqn}
y_{eff} \equiv \frac{Z_{gas}}{\ln{(1/f_{gas})}},
\end{equation}
which will be constant ($y_{eff}=y_{true}$) for any galaxy that has
evolved as a closed-box, assuming that the nucleosynthetic yield is
invariant.  In contrast, if any gas has either entered or left the
galaxy, the measured effective yield will drop\footnote{Note that there
is no plausible process that can increase the effective yield, as
proved by \citet{edmunds90} and shown graphically in an elegant paper
by \citet{koppen99}.  The only exception is accretion of metal-rich
gas -- a highly unlikely possibility that will not be considered
further in this paper.  Thus, the largest observed value of $y_{eff}$
is a lower limit to the true nucleosynthetic yield.} below the closed-box
value of $y_{eff}=y_{true}$, due to changes in the metallicity
$Z_{gas}$ and/or the gas mass fraction $f_{gas}$.  The effective
yield is therefore an observationally determined quantity that can be
used to diagnose departures from closed-box evolution.

Unfortunately, low values of $y_{eff}$ alone do not immediately reveal
{\emph{why}} a system has departed from closed-box evolution.  Either
the addition of metal-poor gas, or the removal metal-rich gas could
suppress the measured effective yield.  It is therefore premature to
assume that supernova-driven outflows alone can explain the low
effective yields seen for galaxies with $V_c\lesssim 100-120\kms$.

This paper explores ways to distinguish between infall and outflow
using the effective yield and the gas mass fraction.
\S\ref{effectiveyieldobssec} summarizes the current observations of
effective yields.  Calculations in \S\ref{gasflowsec} show how infall
(\S\ref{infallsec}), outflow (\S\ref{outflowsec}), and the subsequent
return to closed-box evolution (\S\ref{evolutionsec}) change the
effective yield and the gas mass fraction. A comparison with observational
data in \S\ref{pilyuginsec} leaves metal-rich outflows as the only
viable mechanism for producing the low effective yields observed in
gas-rich galaxies. \S\ref{thresholdsec} stresses the importance of
high gas-richness in allowing enriched outflows to suppress the
effective yield and suggests that the observed mass-dependent
threshold in the effective yield is more closely linked to low star
formation efficiencies than to the depth of the potential well.  Simple
models in \S\ref{modelsec} derive the dependence of mass and
metal-loss as a function of galaxy rotation speed.  Following the
conclusions in \S\ref{conclusionsec}, Appendix~\ref{proofsec} presents
a proof showing how the effective yields calculated for simple
``impulsive'' gas flows are limiting cases of those that result from
arbitrary chemical evolution histories.  Appendix~\ref{garnettsec}
re-examines the conclusions of the paper using the smaller sample of
abundances calculated by \citet{garnett02}.

\section{Observations of the Effective Yield}  \label{effectiveyieldobssec}

\subsection{The Data}

The analysis in this paper is based upon the compilation of effective
yields and gas mass fractions from \citet{pilyugin04}.  This data set
contains the largest number of spirals and dwarf irregulars with uniformly
derived abundances, gas masses, and stellar masses.  It is significantly
larger than the \citet{garnett02} sample, which is analyzed separately
in Appendix A, and has more accurate measurements of the gas content
than \citet{tremonti04}.

The effective yields in \citet{pilyugin04} were calculated using gas-phase
oxygen abundances measured from HII region spectroscopy using the
``$P$-method'', developed in \citet{pilyugin00,pilyugin01a} to match
accurate oxygen abundances based on temperature-sensitive line ratios
involving the weak [{\sc{OIII}}]$\lambda$4363 line (``$T_e$-method'').
The $P$-method was used to derive abundances in high metallicity
($12+\log_{10}{O/H}>8.2$) ``upper-branch'' HII regions of spirals, and
results in systematically lower metallicities than the more widely
adopted $R_{23}$ method \citep{pagel79}.  The metallicities of
irregulars compiled in \citet{pilyugin04} were derived by
\citet{richer95} using the $T_e$-method, or by \citet{pilyugin01b}
using the $P$-method.  All metallicities were interpolated from the
observed radial metallicity gradients to a common galactocentric
radius of $0.4R_{25}$.  This fiducial radius will not occur at the
same number of disk scale lengths in all galaxies, due to variations
in galaxy surface brightness.  Since lower mass galaxies have lower
surface brightnesses on average, their metallicities will be biased
toward smaller radii and higher metallicities than the high mass
galaxies in the \citet{pilyugin04} sample.

The gas mass fractions for the \citet{pilyugin04} sample were
calculated using both the molecular and atomic components measured
primarily from single dish observations, assuming an N(H$_2$)/I(CO)
conversion factor of $1\times10^{20}\cm^{-2}(\K\kms)^{-1}$.  I have
scaled the data to include an additional correction for the mass in
Helium that had been neglected in \citet{pilyugin04}'s Tables~5~\&~7.
The stellar masses were derived assuming a constant $B$-band stellar
mass-to-light ratio of $M/L=1.5$ for the spirals and $M/L=1$ for the
irregulars.  

\subsection{Observed Trends of $y_{eff}$, $f_{gas}$, and $V_c$}

\placefigure{pilyugindatafig}

Figure~\ref{pilyugindatafig} shows the relationships among $y_{eff}$,
$f_{gas}$, and $V_c$ for galaxy disks, using data from
\citet{pilyugin04}'s Tables~5~\&~7.  The left and central panel shows
how the effective yield varies with galaxy rotation speed (left) and
gas mass fraction (center).  Spiral disks are plotted as solid points
(including only types Sbc and later) and irregulars as open circles.
The rightmost panel shows the relationship between $f_{gas}$ and
$V_c$, which will be used in later sections.

The effective yield data in Figure~\ref{pilyugindatafig} show three
main trends.  First, the effective yield increases with the dynamical
mass of a galaxy, as measured by its rotation speed.  As discussed in
\S\ref{introsec}, this trend is typically interpreted as evidence for
larger gas outflows in lower mass galaxies.  Second, the effective
yield saturates at $\log{_{10}(y_{eff})}\approx -2.4$.  I will adopt
this as the true nucleosynthetic yield (i.e.\ $y_{true}=0.004$)
throughout this paper\footnote{Models in \S\ref{modelsec} find that
  true yields in the range $\log{_{10}(y_{true})}\approx -2.5$ to
  $-2.3$ also provide statistically equivalent fits to the data.}.
This yield is within the range of theoretical determinations of the
oxygen yield (see compilation in \citealt{henry00} for a Salpeter
initial mass function), although it is perhaps on the low side.  Third,
the galaxies with low effective yields are relatively
gas-rich\footnote{Note, however, a recent paper by \citet{vanzee06}
  that finds the opposite trend within a sample of isolated dwarf
  irregulars.  There is currently no satisfactory way to reconcile
  these two contradictory results.  Possible, but unattractive
  solutions are: (1) that the behavior seen by \citet{vanzee06} is
  confined only to dwarf irregulars; or (2) that uncertainties in the
  gas mass fraction are partially responsible, given that errors in
  $f_{gas}$ scatter points along the observed \citet{vanzee06}
  relation.}.

\subsection{Uncertainties}

One uncertainty in the data shown in Figure~\ref{pilyugindatafig} is
the appropriate value of $f_{gas}$.  The equations of chemical
evolution track a system where all of the gas is affected by
enrichment.  However, the existence of radial abundance gradients
\citep[e.g.\ ][]{zaritsky94,vanzee98} implies that the entire gas
reservoir of a disk galaxy does not necessarily participate equally in
the chemical evolution of the system.  Thus, the gas that is enriched
may only be a fraction of the total gas mass.  This discrepancy may be
particularly severe for dwarf irregulars, whose HI envelopes extend
well beyond the optical radius of the galaxy.  

To constrain the impact of the uncertainty in the appropriate gas
mass, the small dashed lines in Figure~\ref{pilyugindatafig} show the
locus of the effective yield and gas mass fraction if the appropriate
gas mass is up to a factor of two times smaller than assumed.  This
correction reduces the effective yield, but not significantly.  The
small solid line is the equivalent locus if the stellar mass has been
overestimated by up to a factor of two, due to uncertainties in the
appropriate stellar mass-to-light ratio.  In general, these
uncertainties are much smaller than the range of $y_{eff}$ and
$f_{gas}$ spanned by the data, and could not erase the trends seen in
Figure~\ref{pilyugindatafig}.  The uncertainty in $f_{gas}$ does not
affect the metallicity $Z_{gas}$ in Equation~\ref{peffeqn}, because
the metallicity is determined locally within an HII region through an
analysis of emission lines, and does not require knowledge of the
total gas reservoir.

\section{How Gas Flows Change the Effective Yield}   \label{gasflowsec} 

If the true nucleosynthetic yield is roughly constant among galaxies,
then the observational data in Figure~\ref{pilyugindatafig} indicate
that gas flows must have reduced the effective yield.  Before
performing detailed calculations, it is useful to explore how gas
flows change the effective yield of gas-rich systems, which are the
only ones observed with very low effective yields.

First, expanding the definition of the effective yield
(Equation~\ref{peffeqn}) in terms of the gas mass $M_{gas}$, the
stellar mass $M_{stars}$, and the mass $M_Z$ in metals in the gas
phase, gives
\begin{equation}    \label{peff2eqn}
y_{eff} \equiv \frac{M_Z}{M_{gas}}
         \left[ \frac{1}
                    {\ln{\left(\frac{M_{gas} + M_{stars}}{M_{gas}}\right)}}
              \right].
\end{equation}
\noindent Although $y_{eff}$ was defined based on the closed-box
model, it can be calculated for any system, even if $M_{gas}$,
$M_{stars}$, and $M_Z$ are not related to each other according to a
closed-box model.  Equation~\ref{peff2eqn} therefore holds for any
past star formation and gas accretion/outflow history.

When the gas mass is much larger than the stellar mass, as seen in low
mass late-type galaxies with low effective yields, a Taylor series
expansion of Equation~\ref{peff2eqn} yields
\begin{equation}    \label{peff_highfg_eqn}
y_{eff} \approx \frac{M_Z}{M_{stars}}
\end{equation}
\noindent At high gas mass fractions, the effective yield is therefore
independent of how much gas is in the system.  This result immediately
suggests that accreting even large amounts of gas will make little
change in the effective yield of a system that is already gas-rich.
Although the additional gas will indeed drop the metallicity of the
system (assuming the new gas is metal-poor), it will also increase the
gas richness, and thus keep the system near the relationship between
$Z_{gas}$ and $f_{gas}$ expected for a closed-box system.

Equation~\ref{peff_highfg_eqn} also suggests that metal-rich outflows
will have a significant impact on the effective yield in gas-rich
systems.  The effective yield is linear with the mass of metals in the
gas phase for a gas-rich system, and thus any process which removes
metals leads to an immediate drop in the effective yield.

The gas-rich limit explored above agrees with the more detailed
calculations below, which show that the response of $y_{eff}$ to gas
flows is generally insensitive to both inflows (\S\ref{infallsec}) and
unenriched outflows (\S\ref{outflowsec}), but is extremely sensitive
to metal-rich outflows (\S\ref{outflowsec}).

\subsection{The Response of $y_{eff}$ to Infall}  \label{minimumsec}\label{ineffectivesec}\label{infallsec}

First consider the extreme case where gas is added to a system, but no
additional stars or metals are formed in response.  This scenario
produces the maximum possible decrease in the effective yield for a
given amount of accretion, as shown in Appendix~\ref{proofsec},
assuming the same initial and final gas mass fractions.  Assuming that
a system starts with an initial effective yield $y_{eff,initial}$, gas
fraction $f_{gas,initial}$, gas mass $M_{gas}$, and metallicity
$Z_{gas}$ and then accretes $\Delta M_{gas}$ of gas with
metallicity $Z_{infall}$, the effective yield observed immediately
after gas accretion is
\begin{equation}             \label{ratiofgeqn}
\frac{y_{eff}}{y_{eff,initial}} = 
               \left[\frac{ 1 + \left(\frac{\Delta M_{gas}}{M_{gas}}\right)
                           \left(\frac{Z_{infall}}{Z_{gas}}\right)}
                          { 1 + \left(\frac{\Delta M_{gas}}{M_{gas}}\right)}
                     \right]
               \left[\frac{\ln{\left(f_{gas,initial}\right)}}
                          {\ln{\left(f_{gas,final}\right)}}\right]
\end{equation}
\noindent where
\begin{equation}                \label{fgaseqn}
f_{gas,final} = f_{gas,initial} 
               \left[ \frac{1 + \left(\frac{\Delta M_{gas}}{M_{gas}}\right)}
             {1 + f_{gas,initial}\left(\frac{\Delta M_{gas}}{M_{gas}}\right)} 
                 \right].
\end{equation}
\noindent The ratio of the final to initial effective yield therefore
depends only on the initial gas mass fraction $f_{gas,initial}$, the
ratio $\Delta M_{gas}/M_{gas}$ between the mass of accreted gas and
the initial gas mass, and the ratio $Z_{infall}/Z_{gas}$ between
the metallicity of the infalling gas and that of the initial gas
reservoir.  

The left panel of Figure~\ref{yieldratfig} shows the ratio
$y_{eff}/y_{eff,initial}$ for several values of the initial gas mass
fraction, assuming that $Z_{infall}=0$.  A few clear trends are
apparent.  First, the effective yield falls as more gas is added to
the system.  Second, even when a large amount of gas is added, the
effective yield does not drop to an arbitrarily low value and instead
levels out to a minimum.  Finally, the value of the minimum effective
yield depends on the initial gas richness, with smaller changes in
$y_{eff}$ produced in systems that were initially gas-rich.  

\placefigure{yieldratfig}

For a given initial gas fraction,
Equation~\ref{ratiofgeqn} reaches a minimum as $\Delta
M_{gas}/M_{gas}\rightarrow\infty$ when $Z_{infall}=0$.  In this limit,
the ratio of the minimum to the initial effective yield becomes
\begin{equation}                 \label{mineqn}
\left.\frac{y_{eff}}{y_{eff,initial}}\right|_{min} = 
                \ln{(1/f_{gas,initial})} 
		\left[ \frac{f_{gas,initial}}{1-f_{gas,initial}} \right].
\end{equation}
\noindent This limit uses the fact that $1+\Delta
M_{gas}/M_{gas}=(f_{gas,final}/f_{gas,initial})
(1-f_{gas,initial})(1-f_{gas,final})^{-1}$ and is plotted in
Figure~\ref{ratiominimumfig} as a function of $f_{gas,initial}$.  As
expected from Figure~\ref{yieldratfig}, the effective yield can only
be suppressed by $\sim$30-50\% for the most gas-rich galaxies
\citep[$f_{gas}\sim0.6$; e.g.\ ][]{west05}.

\placefigure{ratiominimumfig}

The response of a galaxy to gas inflow as a function of the initial
gas richness is shown in the left panel of Figure~\ref{yieldfgasfig},
assuming an initial effective yield of $y_{eff}=y_{true}$.  Each solid
line shows the increase in gas richness and the decrease in effective
yield expected for an instantaneous doubling of the galaxies' gas mass
due to accretion of metal-free gas.  As expected, the drop in
effective yield is largest for the most gas-poor galaxies.  However,
in no case is the drop in effective yield greater than 60\%, even for
the rather extreme gas accretion shown.  In contrast, the data in
Figure~\ref{pilyugindatafig} show nearly a factor of ten range of
effective yield, immediately suggesting that infall alone cannot be
responsible for the low effective yields observed.
\S\ref{pilyuginsec} discusses this comparison in more detail.

\placefigure{yieldfgasfig}

\subsection{The Response of $y_{eff}$ to Outflow} \label{outflowsec}  \label{outflowmathsec}

To calculate the response of the effective yield to outflow, consider
an ``impulsive'' gas loss event, i.e.\ one that is not interleaved
with any star formation or gas accretion, similar to the infall case
considered in \S\ref{infallsec}.  This case is a close analog to what
might be expected for a wind driven by a cluster of Type II supernovae
that followed a burst of star formation; for details of the wind
mechanism, see recent reviews by \citet{veilleux05} and
\citet{martin04}.  As shown in Appendix~\ref{proofsec}, the effective
yield immediately following an impulsive outflow will always be larger
than the effective yield that would result from a more general,
continuous outflow driven by on-going star formation, assuming that
both cases have the same initial and final gas mass fractions, and the
same mass in the outflow.

To calculate the impulsive outflow's impact on the effective yield,
first define the mass of gas lost due to outflow as $\Delta M_{gas}$
and the mass of metals lost as $\Delta M_Z$.  The metallicity of the
outflow $Z_{outflow}$ is then
\begin{equation}
Z_{outflow} = \frac{\Delta M_Z}{\Delta M_{gas}}.
\end{equation}
\noindent The outflow metallicity can be parameterized as a multiple
$x$ of the gas phase metallicity at the time of outflow
\begin{equation}
x \equiv Z_{outflow} / Z_{gas}. 
\end{equation}
The maximum value of $x$ is $x_{max}=M_{gas}/\Delta M_{gas}$, which
corresponds to a complete loss of metals ($\Delta M_Z = M_Z$).  

The ratio $x$ can be constrained observationally using X-ray
spectroscopy of hot gas above the midplane of starburst galaxies
combined with abundance analyses of HII region emission lines.
However, $x$ can also be linked to theoretically motivated quantities
by expressing it in terms of the mass fraction of the wind that is
entrained gas from the interstellar medium (ISM).  In terms of this
entrainment fraction $\epsilon$,
\begin{equation}            \label{entraineqn}
x=\epsilon+\left(\frac{Z_{SN}}{Z_{gas}}\right)(1-\epsilon)
\end{equation}
where the metallicity of the supernova ejecta is $Z_{SN}$.  The
entrainment fraction can be also expressed in terms of the more common
``mass loading factor'' $\chi$ as $\epsilon=\chi (\chi+1)^{-1}$.

With the above definitions, the effective yield of a
galaxy with initial effective yield $y_{eff,initial}$ and gas mass
fraction $f_{gas,initial}$ becomes
\begin{equation}                         \label{ratiofgoutfloweqn}
\frac{y_{eff}}{y_{eff,initial}} = 
      \left[\frac{1-x\frac{\Delta M_{gas}}{M_{gas}}}
                 {1-\frac{\Delta M_{gas}}{M_{gas}}} \right]
      \left[\frac{\ln{(f_{gas,initial})}}
                 {\ln{(f_{gas,final})}} \right]
\end{equation}
\noindent immediately after outflow, where the final gas mass fraction
$f_{gas,final}$ is
\begin{equation}
f_{gas,final} = f_{gas,initial} \left[
       \frac{1-\frac{\Delta M_{gas}}{M_{gas}}}
            {1-f_{gas,initial}\frac{\Delta M_{gas}}{M_{gas}}} \right].
\end{equation}
The ratio of the final to initial effective yield therefore depends on
the initial gas mass fraction $f_{gas,initial}$, the fraction $\Delta
M_{gas}/M_{gas}$ of the initial gas mass lost to outflow, and the
ratio $x$ of the metallicity of the outflow to the initial metallicity
of the gas.  

\subsubsection{The Value of $x$ for Realistic Outflows}

Although $x$ can in principle take any value, there are two cases
which bracket most realistic supernova- or AGN-driven outflows.  In
the first, the gas in the outflow is dominated by material originally
surrounding the supernovae that was driven out after acceleration by
shocks.  This ``blast-wave'' outflow will have the metallicity of the
current ISM, and $x\!\approx\! 1$ (corresponding to an entrainment factor
of $\epsilon\!\approx\! 1$).  I will refer to this as the ``unenriched
outflow'' case.  For the second class of wind, the material in the
outflow is dominated by the ejecta from the supernovae driving the
wind.  In this scenario, the outflow will be enriched compared to the
typical ISM, and $x\!\ge\! 1$; I will refer to this as the ``enriched
outflow'' case.  

To derive the value of $x$ for enriched outflows, I first parameterize
the metallicity $Z_{SN}$ of the SN ejecta as a multiple $\eta$ of the
nucleosynthetic yield:
\begin{equation}
Z_{SN}=\frac{M_{Z,SN}}{M_{ejecta}}\equiv \eta y_{true}
\end{equation}
where $M_{Z,SN}$ and $M_{ejecta}$ are the masses in metals and in gas,
respectively, of the SN ejecta.  Using the definition of the effective
yield ($y_{true}\equiv M_{Z,fresh}/M_{remnants}$),
\begin{equation}
\eta \!\equiv\! \frac{M_{remnants}}{M_{ejecta}}\frac{1}{f_{fresh}},
\end{equation}
where $M_{remnants}$ is the mass in long-lived stellar remnants and
$f_{fresh}\!\equiv\! M_{Z,fresh}/M_{Z,SN}$ is the fraction of the
metals in the ejecta that are freshly produced.  For oxygen, SN ejecta
are almost always dominated by fresh production, with $f_{fresh}
\!\gtrsim\!  0.9$ for all but super-solar initial stellar
metallicities \citep[e.g.,][Figure 2]{chieffi04}.  I will therefore
assume $f_{fresh} \!\approx\! 1$.  Note that if
$f_{fresh}\!\approx\!1$, then $\eta\!\approx\!
(1-R_{M>1\Msun})/R_{M>8\Msun}$, where $R$ is the returned mass
fraction for stars evolving in the specified mass range, assuming
that: (1) stars with $M \!<\! 1\msun$ are unevolved and are locked up
over all timescales of interest; (2) only stars with $M \!>\!  8\msun$
(i.e., those with lifetimes $<\! 50\Myr$) contribute material to the
SN ejecta; and (3) only stars with $M \!>\! 8\msun$ produce oxygen.
The value of $\eta$ is therefore insensitive to the details of
explosive nucleosynthesis, and instead depends almost entirely on the
IMF and on the mass lost in SN and stellar winds.

I have calculated $\eta$ for a variety of initial mass functions
(IMFs), using the final remnant masses from \citet[][Table
10]{portinari98} for massive stars ($M_{initial}\!>\!8\msun$) and from
\citet{ferrario05} for intermediate mass stars ($1\msun \!<\!
M_{initial}\!<\!8\msun$).  The resulting values of $\eta$ are $\eta
\!=\! 4.5-5.2$ for the \citet{kroupa01} IMF, $\eta\!=\!6.2-7.1$ for
the \citet{salpeter55} IMF, $\eta\!=\!9.2-10.2$ for the
\citet{kroupa93} IMF, and $\eta\!=\!16.8-18.6$ for the \citet{scalo86}
IMF, assuming stellar metallicities between $z\!=\!0.004-0.02$.  These
variations are driven primarily by differences in the high mass end of
the IMF, as can be seen by considering the simplified case where all
stars with $>\!1\msun$ are completely disrupted either through stellar
winds or SN.  In this case, $\eta \!\approx\!
f_{mass}(M\!<\!1\msun)/f_{mass}(M\!>\!8\msun)$ where $f_{mass}$ is the
mass fraction of stars in the specified mass range for a given IMF.
This estimate gives values of $\eta$ that are $\sim$30\% lower than
the full calculation.  The fraction of mass in low mass stars varies
by $\sim$15\% ($f_{mass}(M\!<\!1\msun)\!=\!0.57- 0.67$) for the
various IMFs considered, but the mass fraction of high mass stars
varies by a factor of three ($f_{mass}(M\!>\!8\msun)\!=\!0.05-0.17$,
from the Salpeter to the Kroupa 2001 IMFs, respectively).  The
variation in $f_{mass}(M\!>\!8\msun)$ therefore drives the majority of
the variation in $\eta$ among different IMFs, more so than differences
in the assumed metallicity or mass-loss model.

Current data tend to favor the shallower top end slope adopted by
\citet{kroupa01} \citep[see review by][]{chabrier03}, and thus I will
adopt a value of $\eta\approx 5$ for the remainder of this paper.  If
the true value of $\eta$ is higher, then larger entrainment factors
would be needed to match the same outflow metallicity.  Note that even
with this lower value of $\eta$, the wind will be dominated by freshly
produced metals for any reasonable value of $f_{gas}$; only a system
with $f_{gas}\!<\!\exp{(-\eta)}\!=\!0.007$ would be sufficiently metal-rich
for the returned metals to equal the fresh production, assuming that
the stars driving the wind were formed recently (i.e.\ as for a Type
II supernova-driven wind) and have metallicities approximately equal
the current gas-phase metallicity.  Thus $x\!=\!\eta y_{true} / Z_{gas}$
will be greater than 1 for all reasonable gas fractions.

\subsubsection{Results for Outflows}

Using Equation~\ref{ratiofgoutfloweqn} with the appropriate values of
$x$, Figure~\ref{yieldratfig} plots the ratio of final to initial
effective yield as a function of the mass fraction of gas lost in the
outflow ($\Delta M_{gas}/M_{gas}$) for an unenriched outflow (center
panel) and for a maximally enriched outflow with $\epsilon\!=\!0$ (right
panel).

For unenriched outflows, there are two main results.  First,
unenriched outflows are most effective in reducing the effective yield
of gas-rich galaxies.  This trend is in contrast to infall, which
causes the largest reductions in $y_{eff}$ when galaxies are gas-poor.
Second, even for the most gas-rich galaxies, reducing the effective
yield by a factor of ten requires nearly complete removal of the ISM.
Thus, the effective yield is relatively insensitive to outflows that
drive out the existing ISM, except in the most extreme case of large
gas losses ($\gtrsim$75\%) from very gas-rich systems.

For enriched outflows, the situation is quite different.  The
effective yield is extremely sensitive to gas loss from gas-rich
galaxies, and decreases without limit for even modest gas loss
($\lesssim$10\%). 

The sensitivity of $y_{eff}$ to outflow can also be seen in the
central and right hand panels of Figure~\ref{yieldfgasfig}, for
unenriched and enriched outflows, respectively.  For galaxies with an
initial effective yield of $y_{eff}=y_{true}$ and a range of initial
gas mass fractions, these plots show how outflows reduce both the
effective yield and the gas richness.  Similar to the infall models
shown in the left panel, unenriched outflows produce only modest
changes in the effective yield, even when half the gas is lost from
the system.  In contrast, enriched outflows easily produce dramatic
drops in the effective yield, particularly for larger gas mass
fractions.  Metal-enriched outflows are therefore the
only viable mechanism for producing the extremely low effective yields
seen in Figure~\ref{pilyugindatafig}.  \S\ref{pilyuginsec} will
discuss this point in more detail.

\subsection{Evolution after Gas Accretion or Gas Loss} \label{evolutionsec}

Gas accretion and gas loss are likely to be episodic processes
\citep[e.g.,][]{marlowe95}.  Thus, while they may temporarily reduce
the effective yield, they will likely be followed by
periods of closed-box chemical evolution that alter the effective
yield while decreasing the gas mass fraction.

To calculate the impact of any subsequent closed-box evolution,
assume that a galaxy has an initial effective yield of
\begin{equation}                        \label{yeffpostfloweqn}
y_{eff,postflow} = \frac{Z_{postflow}}{\ln{(1/f_{gas,postflow})}}
\end{equation}
\noindent immediately after gas accretion or gas loss.  The gas mass
fraction before the system returns to closed-box evolution is defined
$f_{gas,postflow} = M_{gas,postflow} / (M_{gas,postflow} +
M_{stars,postflow})$.  When the galaxy returns to evolving as a
closed-box, it will obey the equation for instantaneous recycling of a
system with fixed baryonic mass:
\begin{equation}                \label{dZdMeqn}
\frac{{\rm d}Z_{gas}}{{\rm d}M_{gas}} = -\frac{y_{true}}{M_{gas}},
\end{equation}
\noindent which shows that the gas-phase metallicity increases
proportionally to the true nucleosynthetic yield as the gas mass
decreases.  This equation can be integrated using the metallicity and
the gas mass immediately after infall/outflow as the initial state:
\begin{equation}
\int^{Z_{gas}}_{Z_{postflow}}{\frac{{\rm d}Z_{gas}}{y_{true}}} =
     -\int^{M_{gas}}_{M_{gas,postflow}}{\frac{{\rm d}M_{gas}}{M_{gas}}}
\end{equation}
\noindent yielding
\begin{equation}                 \label{Zpostfloweqn}
\frac{Z_{gas}-Z_{postflow}}{y_{true}} = 
              \ln{\left(\frac{M_{gas,postflow}}{M_{gas}}\right)}.
\end{equation}
\noindent Equation~\ref{Zpostfloweqn} can be rearranged to solve for
$Z_{gas}$ and then substituted into the definition of the effective
yield.  Using the definition of the effective yield to substitute
for $Z_{postflow}$ and the fact that $M_{gas}+M_{stars} =
M_{gas,postflow}+M_{stars,postflow}$ (i.e.\ no infall or outflow), the
effective yield $y_{eff}$ after subsequent closed-box evolution then
becomes
\begin{equation}                        \label{yeffpostfloweqn2}
y_{eff} = y_{true} \left[ 1 - 
               \frac{\ln{(f_{gas,postflow})}}{\ln{(f_{gas})}}
               \left(1 - \frac{y_{eff,postflow}}{y_{true}}\right)  \right],
\end{equation}
where $f_{gas}$ is the gas mass fraction after the system has evolved.
Equation~\ref{yeffpostfloweqn2} shows that when a galaxy returns to
closed-box evolution its effective yield will increase back to the
true nucleosynthetic yield as the gas mass fraction decreases.  Thus,
if star formation is on-going, the reduction of the effective yield is
temporary. This result has also been shown previously by
\citet{koppen05} in a more detailed study of nitrogen and oxygen
abundance ratios during episodic infall.  The return to closed-box
evolution can also be seen in the chemical evolution models of
\citet{pilyugin98}.

To directly track the evolution of the effective yield and gas mass
fraction, Equation~\ref{yeffpostfloweqn2} can be rewritten
\begin{equation}                        \label{yeffpostfloweqn3}
\frac{y_{eff}}{y_{eff,postflow}}  = 1 +
               \frac{\ln{Q}}{\ln{(Q f_{gas,postflow})}}
               \left(\frac{y_{true}}{y_{eff,postflow}} - 1\right),
\end{equation}
where $Q\equiv f_{gas} / f_{gas,postflow}$, a quantity that goes from
1 to 0 as star formation proceeds.  Equation~\ref{yeffpostfloweqn3}
shows that the effective yield measured some time after the end of
infall/outflow will always increase, by an amount that depends on: (1)
how low the effective yield was after the flow stopped; (2) how
gas-rich the galaxy was; and (3) how much the gas fraction has dropped
due to subsequent star formation.

Figure~\ref{evolutionfig} shows the change in $y_{eff}$ and $f_{gas}$
predicted by Equation~\ref{yeffpostfloweqn3}, as a function of the
initial gas mass fraction (light to dark lines), and the initial
postflow effective yield (left, center, and right panels).  These
curves trace non-intersecting streamlines in the plane of $f_{gas}$
and $y_{eff}$.  The curves show the behavior expected from
Equation~\ref{yeffpostfloweqn2}, namely, that as gas converts into
stars and the gas mass fraction decreases, the effective yield rises
back toward the true nucleosynthetic yield expected for a closed-box
model.

\placefigure{evolutionfig}

At very low effective yields, the return to closed-box evolution
produces a steep fractional rise in the effective yield, due to the
linear increase in metallicity with star formation
(Equation~\ref{dZdMeqn}).  Slight decreases in the gas mass fraction
due to star formation can drive the effective yield back up to large
values.  Thus, galaxies will have difficulty maintaining a very low
effective yield after infall or outflow has stopped.  For example, if
a galaxy has an effective yield of ${\rm log}_{10}
y_{eff,postflow}=-3.5$ and gas fraction of $f_{gas,postflow}=0.5$,
then converting just 25\% of its gas into stars increases its
effective yield by nearly a factor of ten.  Figure~\ref{evolutionfig}
therefore suggests an additional obstacle to producing a population of
galaxies with very low effective yields.  These systems must either
experience continual infall or outflow, or they must have highly
inefficient star formation to keep their gas fractions nearly constant
and their effective yields low.  I return to this point in
\S\ref{thresholdsec}.

Figure~\ref{evolutionfig} also helps to explain why massive spiral
galaxies currently have high effective yields.  Spiral disks were
probably once gas-rich \citep{robertson05,springel05}, and thus must
have once been more susceptible to reductions in their effective
yields by metal-enriched winds than they are today.  Indeed,
observations suggest that large scale outflows were common in massive
galaxies at early times \citep[e.g.\ ][and the recent review by
\citealt*{veilleux05}]{adelberger03,steidel04}, and yet the effective
yields of the likely descendants are high.  Figure~\ref{evolutionfig}
and Equation~\ref{yeffpostfloweqn2} show that any reduction in the
effective yield of a gas-rich precursor of a present day spiral must
be temporary.  As the gas-rich disk evolves to its present gas-poor
state, its effective yield will rapidly increase back to the
nucleosynthetic yield.  Thus, the current effective yields of spiral
disks place only limited constraints on their past gas loss, unless
one utilizes the limiting cases provided in Appendix A.  However,
{\emph{in situ}} measurements of $y_{eff}$ at high redshift
\citep[e.g.][]{erb06} could reveal evidence for an earlier period of
outflow.  At these redshifts, spiral galaxies are much more likely to
be gas rich, and thus will be more likely to show low values of the
effective yield for a given amount of outflow.
 
\section{Confronting Infall \& Outflow Models with Data}  \label{pilyuginsec}

A comparison of the models shown in Figure~\ref{yieldfgasfig} to the
central panel of Figure~\ref{pilyugindatafig} immediately suggests
that neither gas accretion nor unenriched outflows can reduce
effective yields to the low levels seen in low mass gas-rich dwarf
irregular galaxies.  Metal-enriched outflows are the only mechanism
apparently capable of sufficiently reducing the effective yields.
These conclusions are strengthened by factoring in the tendency of
post-flow evolution to return galaxies to the true nucleosynthetic
yield.  I now compare each of the three gas flow models with the data
in more detail.  Appendix~\ref{garnettsec} contains an identical
comparison with the earlier data from \citet{garnett02}, reaching
similar conclusions.

\subsection{Infall Models}  \label{infallgarnettsec}

The calculations presented in \S\ref{infallsec} show that there is an
absolute minimum value of the effective yield that can be produced by
gas accretion (Equation~\ref{mineqn}; Figure~\ref{ratiominimumfig}).
Gas-rich galaxies have the largest value of this minimum, making their
effective yields essentially impossible to change by gas accretion.
The existence of gas-rich galaxies with low effective yields therefore
immediately proves that infall alone cannot produce the necessary
reduction in $y_{eff}$.

To show this limit, the left panel of Figure~\ref{pilyuginfig}
reproduces the data points from Figure~\ref{pilyugindatafig},
superimposed with the minimum effective yield that can be reached via
infall (Equation~\ref{mineqn}) at each gas fraction, assuming an
initial effective yield of $y_{eff}=y_{true}$.  Clearly, all of the
irregular galaxies and a few of the spirals (NGC~598, NGC~925, and
NGC~2403) have effective yields that are too low to be explained
solely by infall.  Thus, outflows must have occurred during these
galaxies' evolution.

\placefigure{pilyuginfig}

To demonstrate the limits of infall more clearly, the dotted loci in
Figure~\ref{pilyuginfig} indicate how past
gas accretion could have brought galaxies to their present values of
$y_{eff}$ and $f_{gas}$.  These loci are based on the rather extreme
assumption that the galaxies' gas masses have doubled, and that the
galaxies are being observed immediately after accretion, before any
star formation has taken place.  Along these dotted lines the
effective yield drops and the gas mass fraction increases as gas is
added to a galaxy, moving the galaxy down to lower effective yields
and rightwards toward higher gas mass fractions.  However, even with
these generous assumptions, none of these galaxies' loci intercept the
horizontal dashed line expected for closed-box evolution.  Thus,
before the hypothetical gas accretion, some other process would have 
needed to suppress the galaxies' effective yields below the expected
closed-box value.  Infall alone is therefore incapable of producing the
lowest effective yields.  

The only conceivable way that infall could produce gas-rich galaxies
with effective yields below the minimum predicted by
Equation~\ref{mineqn} is if the initial gas mass fraction were low and
the amount of accreted gas large.  However the data suggests that this
possibility is highly unlikely, given that the required initial gas
fractions are much lower than seen in any disk galaxy.  For example,
for infall to produce galaxies with ${\rm log}_{10} y_{eff}=-3.0$, the
galaxies initially must have had $f_{gas}<0.025$, which is typical of
gas fractions in elliptical galaxies, not disks.  Moreover, a galaxy
with such a low initial gas mass fraction would have a very high
stellar metallicity, which conflicts with the low stellar
metallicities derived for dwarf galaxies using broad-band colors and
resolved stellar populations \citep[e.g.\
][]{macarthur04,skillman03,bell00,holtzman00,vanzee97a}; this would
also lead to a large metallicity difference between the gas phase and
the stars (although see \citealt{venn03} and \citealt{lee05} for a
possible example of such an offset in the dwarf galaxy WLM).  Finally,
the necessary accretion would have to be extremely large.  For
accretion to bring a galaxy with an initial gas fraction of
$f_{gas}\!<\!0.025$ up to a final gas fraction of $f_{gas}\!=\!0.5$, the
galaxy's gas mass would have to increase by a factor of nearly twenty,
and its baryonic mass would have to double.  However, even this large
amount of accretion would not lower the effective yield by the factor
of ten needed to explain the lowest mass galaxies.  Infall onto a
gas-poor system is also unlikely to produce low effective yields in
disks that ``regrow'' around gas-poor bulges, lahtough they do meet
the condition for significant gas accretion onto gas-poor systems;
subsequent star formation in the gaseous disk would have wiped out any
temporary reduction in the effective yield, as shown in
\S\ref{evolutionsec}.

Although the comparisons in Figure~\ref{pilyuginfig} prove that infall
alone is not responsible for low effective yields, they do not
necessarily imply that infall has not occurred.  Instead, the opposite
is true.  Because the effective yield is insensitive to infall, gas
accretion leaves little trace on the effective yield, particularly for
gas-rich galaxies.  The effective yields of gas-poor galaxies are
potentially more sensitive to past infall (e.g.\ see the evolution
loci in the upper left of Figure~\ref{pilyuginfig}).  However, the
spiral galaxies which dominate the gas-poor population of disks are
also observed to have high star formation efficiencies \citep[e.g.\
][]{kennicutt98b}, and thus should rapidly consume any accreted gas,
driving their effective yields back up toward the nucleosynthetic
yield (Figure~\ref{evolutionfig}).  The effective yields of both
gas-rich and gas-poor disk galaxies are therefore unlikely to show
{\emph{any}} sign of past infall, even if it has occurred.

Given that infall has been largely overlooked in recent years due to
the pervasive theoretical focus on outflows, it is worth reiterating
that the evidence for significant sustained infall of onto disk
galaxies is substantial, and that infall of
$\sim\!1\msun\pc^{-2}\Gyr^{-1}$ onto high mass galaxies is required
(1) to solve the G-dwarf \citep{vandenbergh62,schmidt63} and K-dwarf
problems \citep{favata97,rochapinto98,kotoneva02}, as originally
suggested by \citet{larson72}; (2) to provide proper relative
abundances of different elements and an extended star formation
history in the Milky Way \citep[e.g.\ ][and many others]{chiappini97,
bossier99, chang99, chiappini01, alibes01, fenner03,casuso04}; (3) to
explain the high deuterium abundances seen in the Milky Way
\citep[e.g.\ ][]{quirk73,chiappini02} and (4) to explain the
broad-band colors, gas-fractions, and metallicities of galaxies at low
and high redshift \citep{boissier00,ferreras04}.  In low mass
galaxies, infall is required to explain why their star formation rates
are typically observed to rise to the present day
\citep[e.g.][]{brinchmann04,gavazzi02}.  Such a systematic rise can
only occur if the gas surface density is rising with time, as would be
expected for gas infall onto a system with inefficient star formation
\citep[as dwarf galaxies are observed to have; e.g.\
][]{vanzee97b,vanzee01,hunter04}.  In total, there is strong ancillary
evidence for on-going gas accretion in both low mass irregular and
high mass spirals, but no evidence can come from observations of the
effective yield alone.

\subsection{Unenriched Outflow Models}  \label{unenrichedoutflowgarnettsec}

The central panel of Figure~\ref{yieldfgasfig} shows that the
effective yield is nearly as insensitive to unenriched ``blast wave''
outflows as to gas accretion.  The only significant difference is that
unenriched outflows have the opposite dependence on the initial gas
mass fraction, causing larger drops in the effective yield for
gas-rich galaxies.  Even at high gas mass fractions, however, the
decrease in effective yield is still far too small to explain the
range of effective yields seen in Figure~\ref{pilyugindatafig}.

The central panel of Figure~\ref{pilyuginfig} shows this limitation
clearly.  The dotted loci show how the galaxies analyzed by
\citet{pilyugin04} could have arrived at their present gas mass
fractions and effective yields by losing 25\% of their ISM in an
outflow, in the optimistic case that all galaxies were observed
immediately after outflow ceased, but before subsequent star
formation drove their effective yields back to larger
values.  If a locus intercepts the horizontal line at the adopted
initial yield of $y_{true}$, then it is possible that unenriched
outflows could have brought the galaxy off the expected closed-box
evolution to its present position.  However, even with these generous
assumptions it is clear that all but one of the dwarf irregulars could
not have had a closed-box effective yield before the hypothetical
outflow started, and that some other process must have already reduced
the effective yield.  

The central panel of Figure~\ref{pilyuginfig} shows that the
effective yields of massive spirals are also unlikely to have been
significantly altered by unenriched outflows. This result is not
surprising, given the weaker response of the effective yield to outflow
from gas-poor systems.

As with the infall case discussed above, the data in
Figure~\ref{pilyuginfig} should not be interpreted as evidence against
unenriched outflows.  Indeed, the calculations in this paper suggest
that a large fraction of a galaxy's ISM can be removed without
significantly decreasing the galaxy's effective yield.  Instead, the
data should be interpreted as evidence that unenriched outflows alone
cannot be the cause of the low effective yields seen in low mass
galaxies.

\subsection{Enriched Outflow Models}  \label{enrichedoutflowgarnettsec}

Having ruled out inflow and unenriched infall in
\S\ref{infallgarnettsec}~\&~\S\ref{unenrichedoutflowgarnettsec},
the only possible mechanism that can significantly lower the effective
yield is metal-enriched outflow, such as would be caused by direct
escape of SN ejecta.  The models in Figure~\ref{yieldfgasfig} show
that the effective yields of gas-rich galaxies are extremely sensitive
to even modest amounts of gas loss, provided that the ejected gas is
metal-rich.  An enriched outflow that removes less than a fifth of a
galaxy's gas can drop the effective yield by more than a factor of
ten, provided that more than half of the galaxy's baryonic mass is
gaseous.

The right panel of Figure~\ref{pilyuginfig} directly compares enriched
outflow models (Equation~\ref{ratiofgoutfloweqn}) with the data.  As
in the other panels, the dotted loci trace how the galaxies could have
reached their present position after driving a metal-enriched wind
that expelled $\lesssim\!5$\% of the galaxies' gas.  Unlike the other
panels, however, the majority of the dotted lines intersect the
closed-box effective yield, indicating that a modest amount of
enriched outflow could have brought the galaxies' effective yields
from the true nucleosynthetic yield down to their present low values.
Moreover, because the outflow models calculated in \S\ref{outflowsec}
give lower limits to the amount of reduction produced by more
continuous outflows, it is possible that even weaker winds may be
sufficient to produce the same reduction in $y_{eff}$ (for example, if
the wind began when the galaxies had higher gas fractions, and then was
followed or interleaved with star formation).

Taken together, the various panels in
Figures~\ref{yieldfgasfig}~\&~\ref{pilyuginfig} explain why there are
few gas-poor galaxies with low effective yields.  There are simply no
viable scenarios capable of producing such systems.  Indeed, all three
mechanisms explored in this paper leave the region with ${\rm
  log}_{10} y_{eff} \lesssim -2.8$ and $f_{gas}\lesssim 0.4$
vacant for any reasonable amount of gas accretion or gas loss.

\subsection{Subsequent Evolution Models}  \label{subsequentevsec}

The dotted loci in Figure~\ref{pilyuginfig} assume that all galaxies
have been observed immediately after infall or outflow.  However,
given the episodic nature of these processes, it is more likely that
most of the galaxies are being observed in a more quiescent state, and
are turning gas into stars without any significant gas flows.  The
galaxies are therefore more likely to have recently evolved as a
closed-box.

\placefigure{pilyuginevolutionfig}

To demonstrate how recent closed-box evolution could have brought the
galaxies to their current effective yields,
Figure~\ref{pilyuginevolutionfig} shows the data from
\citet{pilyugin04} along with dotted loci indicating their past
effective yields and gas mass fractions, assuming that star formation
has reduced their initial gas fraction by 5\% to reach their present
values.

As expected from Figure~\ref{evolutionfig},
Figure~\ref{pilyuginevolutionfig} shows that galaxies with low
effective yields are extremely sensitive to any subsequent star
formation.  If such galaxies have recently converted just a small
fraction of their gas into stars, then their past effective yields
must have been even lower than currently observed.  Maintaining low
effective yields therefore requires either nearly continuous
metal-rich outflows or extremely low star formation rates.

\section{The Break in $y_{eff}$ vs $V_c$: Outflows \& Star Formation Efficiency}  \label{thresholdsec}

\subsection{A Mass Threshold for Gas Loss?}

The onset of low effective yields in low mass galaxies is usually
interpreted as evidence for a mass-dependent threshold for mass loss
by winds \citep[e.g.][]{garnett02,tremonti04}.  In these scenarios,
the fraction of gas and metals that remain bound to a galaxy is
assumed to be a strong function of the depth of the galaxy's
gravitational potential well.  Low mass galaxies have shallow
potentials, and thus should retain only a small fraction of their gas
for winds whose energies excede the depth of the potential.  For
example, a simple scaling argument presented by \citet{tremonti04}
suggested that galaxies with circular velocities of less than $V_c\sim
85\kms$ retain fewer than half of their metals.

In contrast to this widely adopted picture, the results above (and
Figure~\ref{yieldfgasfig} in particular) suggest that massive galaxies
may also have expelled a large fraction of their gas.  Although
these galaxies are observed to have high effective yields, they are
also moderately gas-poor and thus their effective yields can never be
significantly reduced by outflows.  Massive disk galaxies also have
continuous, relatively high rates of star formation, and thus their
effective yields rebound quickly from any temporary reduction.  Thus,
the drop in the effective yield seen in galaxies with $V_c \lesssim
120\kms$ does not necessarily indicate that {\emph{only}} these lower
mass galaxies have experienced outflows. 

If massive galaxies have also driven strong winds, then the trend
towards low effective yields is not necessarily due to increasing mass
loss.  Instead, it can be due to low mass galaxies' higher
gas richness\footnote{A Spearman rank correlation test on the data
from \citet{pilyugin04} shows that the correlation between
$\log{_{10}V_c}$ and $f_{gas}$ is stronger than between either
quantity and $\log{_{10}y_{eff}}$.}, which increases the sensitivity
of the effective yield to mass loss.  Thus, even if all galaxies lost the
{\emph{same}} fraction of gas, the effective yield would be lower in
dwarf galaxies, due to their systematically higher gas fractions.
The actual dependence of outflows on galaxy mass is calculated
below in \S\ref{modelsec}, after including variations due to
gas richness, and is found to be much weaker than previously assumed.

\subsection{Why is There a Break?}

Given that the gas fraction varies smoothly with rotation speed, why
is there an apparent break in the relationship between $y_{eff}$ and
$V_c$?  Are there plausible mechanisms that could produce a transition
at $V_c\!\sim\!120-125\kms$, other than a sharp increase in mass loss?
First, inspection of the left panel of Figure~\ref{yieldfgasfig} shows
that $y_{eff}$ responds non-linearly to the gas mass fraction for a
fixed amount of outflow.  Small changes in $f_{gas}$ produce much
larger changes in $y_{eff}$.  Secondly, several groups
\citep{verde02,dalcanton04} have noted that below
$V_c\!\sim\!120\kms$, disk galaxies have surface densities that lie
entirely below the \citet{kennicutt89} threshold for efficient star
formation.  Their low star formation efficiencies would keep low mass,
low surface density galaxies gas-rich, and would allow low effective
yields to be maintained after they were established.  These two
effects could potentially produce a break in the relationship between
$y_{eff}$ and $V_c$.  By Occam's razor, the coincidence of these two
mass scales is strong circumstantial evidence that the variation in
star formation efficiency, rather than potential well depth, is
critical for producing the transition in the trend of $y_{eff}$ with
$V_c$.

Indeed, only a galaxy that falls off the \citet{kennicutt98b} global
Schmidt law can realistically maintain low effective yields between
episodes of outflow.  To demonstrate the need for low star formation
efficiencies, first assume that outflows are discrete events, separated
temporally by $\Delta t$.  To maintain an effective yield of ${\rm
  log}_{10} y_{eff}<-3$ during the interval between outflows, no more
than 2.5\% of a galaxy's gas ($\Delta f_{gas,SF}<0.025$) must have
converted to stars since the end of the last outflow, setting an upper
limit to the star formation rate.  For a galaxy that follows the
Schmidt law to have a star formation rate less than this upper limit
($<0.25\times10^{-3}\msun\s^{-1}\kpc^{-2}(\Delta t/10^9\yrs)^{-1}
(\Sigma_{gas}/10 \msun\pc^{-2})$), its current gas surface density
must be less than $\Sigma_{gas}<0.0032 \msun/\pc^2 \, (\Delta
f_{gas,SF}/0.025)^{2.5} (\Delta t/10^9\yrs)^{-2.5}$, which is several
orders of magnitudes below what is typically observed.  However, if a
galaxy lies entirely below the \citet{kennicutt89} star formation
threshold, the gas surface density needed to suppress the star
formation rate is much more reasonable.  I have estimated the star
formation law in this regime from Figure~1 of \citet{kennicutt98b} as
$\Sigma_{SFR}\propto\Sigma_{gas}^{12}$, which is drastically steeper
than the Schmidt law.  Assuming that galaxies with
$\Sigma_{gas}\lesssim10\msun/\pc^2$ fall in this low star formation
efficiency regime, they only require that $\Sigma_{gas}<7.5
\msun/\pc^2 \, (\Delta f_{gas,SF}/0.025)^{0.09} (\Delta
t/10^9\yrs)^{-0.09}$ to maintain low effective yields between outflow
events.  Almost all late-type and/or LSB dwarf galaxies have gas
surface densities below this threshold \citep{swaters02}, and thus
they should be capable of maintaining low effective yields in between
episodes of outflow.

\subsection{Possible Limitations}

Given the possibility that the drop in star formation efficiency at
$V_c\sim120\kms$ is responsible for the transition in the effective
yield observed by \citet{garnett02}, it is worth re-examining several
issues.  First is the existence of the transition itself.
\citet{garnett02} based his claim of a transition on a plot of the
logarithm of the effective yield against the rotation speed.  His
Figure~4 showed a flat relationship with rotation speed above
$V_c\!\sim\!125\kms$.  However, when the effective yield is plotted
against the logarithm of the rotation speed, as in
Figures~\ref{pilyugindatafig}~\&~\ref{garnettdatafig}, the evidence
for a sharp transition seems much weaker.  There is little dynamic
range at rotation speeds above $\sim\!100\kms$, such that the data can
be adequately fit by a single power-law, giving $y_{eff}\approx 0.002
\times (V_c/100\kms)^{0.444}$ for the data in
Figure~\ref{garnettdatafig}.  Statistically, however, functions that
have a break in the slope (such as a double power-law) have lower
$\chi^2$ values than expected for the addition of new parameters,
suggesting that there is some change in the behavior of $y_{eff}$ from
low to high galaxy masses.

The other remaining issue is the relative strength of the effective
yield's correlations with gas mass fraction and rotation speed.  If
the gas fraction is more critical to setting $y_{eff}$ than rotation
speed, then $y_{eff}$ should correlate more strongly with $f_{gas}$
than with $V_c$.  The data in Figure~\ref{pilyugindatafig} shows
scatter in $\log{_{10}y_{eff}}$ versus $f_{gas}$ that is 45\% higher
than in the plot of $\log{_{10}y_{eff}}$ versus $\log{_{10}V_c}$,
initially suggesting that a galaxy's effective yield might depend more
strongly on its mass than on its gas mass fraction.  However, the
uncertainty in the measurement of $f_{gas}$ is at least five times
larger than the uncertainty in $V_c$, and produces scatter that is
nearly perpendicular to the mean relationship, as can be seen by the
solid and dashed lines in the right two panels of
Figure~\ref{pilyugindatafig}.  This effect maximally broadens the
relationship between $f_{gas}$ and $y_{eff}$.  The resulting higher
uncertainties leave open the possibility that the actual trend with
gas mass fraction and star formation efficiency is at least as strong
as the trend with rotation speed.

\section{Estimates of Recent Gas Loss}   \label{modelsec}

The previous sections suggest that: (1) only moderate mass loss of
supernova-enriched material is necessary to explain low effective
yields; and (2) that there is unlikely to be a sharp increase in gas
loss below $V_c\sim 100-120\kms$.  These points can be demonstrated
with a simple model that calculates the outflows needed to explain the
observed trends in effective yield.  The model calculates the mass
loss needed for a single outflow event to bring galaxies to their
present gas mass fraction and effective yield, assuming that they
initially followed closed box evolution.  For a fixed entrainment
factor, these assumptions give a strict upper bound to the mass loss
needed to explain the observed effective yields, as shown in the
second lemma of Appendix~\ref{proofsec}; a more realistic extended
outflow and star formation history would lead to effective yields
lower than observed, for any gas loss greater than that calculated
here.

\subsection{Assumptions of the Gas Loss Model}

To estimate the maximum mass lost as a function of galaxy rotation
speed, the current gas mass fraction is parameterized as
$f_{gas}(V_c)\!=\!1.083-0.158\ln{(V_c)}$, with $f_{gas}(V_c)$
constrained to lie between zero and 1.  The effective yield is
parameterized as 
\begin{equation}    \label{yeffparameqn}
y_{eff}(V_c)=y_{true}/[1+(V_0/V_c)^\alpha].
\end{equation}
Values are assigned to the parameters $\alpha$ and $V_0$ using a
$\chi^2$-minimization with respect to the data in
Figure~\ref{pilyugindatafig}, assuming $\log{_{10}(y_{true})}$ between -2.6
and -2.2.  The $\chi^2$ minima for all $\log{_{10}(y_{true})}$ between -2.5 and
-2.3 are statistically equivalent, giving parameters for
Equation~\ref{yeffparameqn} of $\alpha\!=\!1.2$ and $V_0\!=\!31\kms$ for
$\log{_{10}(y_{true})}\!=\!-2.5$, and $\alpha\!=\!0.8$ and $V_0\!=\!80\kms$ for
$\log{_{10}(y_{true})}\!=\!-2.3$.  For these values, plots of
Equation~\ref{yeffparameqn} are nearly indistinguishable over the
range of rotation speeds where data exist.  The best-fit values of
$\alpha\!=\!0.9$ and $V_0\!=\!54\kms$ for $\log{_{10}(y_{true})}\!=\!-2.4$ are
used throughout this section, and are used for the curved line in the
left panel of Figure~\ref{pilyugindatafig}.

The model uses the equations in \S\ref{outflowmathsec} to derive the
rotation speed dependence of the gas loss needed to produce the
observed relations for $f_{gas}(V_c)$ and $y_{eff}(V_c)$.  The outflow
is assumed to consist of $\Delta M_{gas}$ of gas, a fraction
$\epsilon$ of which is entrained gas from the ISM, and $1-\epsilon$ of
which is pure supernova ejecta (Equation~\ref{entraineqn}).  Chandra
observations of dwarf starburst galaxies by \citet{ott05} find
entrainment fractions of $\epsilon\!\sim\!50-83$\%, with the majority
being less than 65\%.  \citet{ott05} argue that these entrainment
factors are probably overestimates of the amount of ISM mixed with the
ejecta, and that the actual entrainment fractions are thus likely to
be smaller.  

As in \S\ref{outflowmathsec}, the metallicity of the supernova ejecta
is assumed to be $Z_{SN}\!=\!\eta y_{true}$, with $\eta\!=\!5$ and
$\log{_{10}(y_{true})}\!=\!-2.4$.  Larger values of $\eta$ increase
the enrichment of the outflow, and reduce the amount of outflow needed
to remove sufficient metals.  Models with higher or lower adopted
values of $y_{true}$ produce similar results for low rotation speeds,
where the differences between the observed effective yields and
$y_{true}$ are large.  The models have larger differences at high
rotation speeds, where the difference between
$\log{_{10}(y_{true})}\!=\!-2.3$ and $\log{_{10}(y_{true})}\!=\!-2.5$
is a large fraction of the difference between the observed effective
yields and $y_{true}$.  However, the resulting uncertainty in the
actual gas and metal-loss at large galaxy masses remains less than a
factor of two.  Theoretical determinations of $y_{true}$ currently
give no help in resolving this issue, given the large spread of
published values, and the strong dependence on the IMF.

\subsection{Results of the Gas Loss Model}

Figure~\ref{modelfig} shows the resulting upper bounds on the fraction
of baryons lost (left panel) and of metals lost (right panel), for a
series of entrainment fractions from $\epsilon\!=\!0$ to $\epsilon\!=\!1$
(i.e.\ from pure SN ejecta to pure blast wave outflows).  

As seen in the left panel, the fraction of the baryons lost rises
monotonically toward low disk masses ($V_c>10\kms$), but shows no
sharp features indicative of a sudden onset of winds.  The maximum
fraction of baryons lost is never more than 15\% for all entrainment
factors less than 75\%, and only reaches above 50\% for
unrealistically high entrainment factors ($\epsilon>94$\%,
corresponding to mass loading of nearly a factor of 20).  There is
also a relatively small range in the baryonic mass fraction lost by
disks.  The range could be as small as a factor of two if
$\log{_{10}(y_{true})}\!=\!-2.3$, and is unlikely to be larger than a
factor of 6-7 (i.e.\ if $\log{_{10}(y_{true})}\!=\!-2.5$).

Given the lack of any obvious feature at $V_c\!\sim\!120\kms$ and the
generally modest overall gas loss implied by Figure~\ref{modelfig},
the idea that winds are dramatically more effective below some
threshold in galaxy mass is not compelling.  Instead, mass loss from
disks seems to be rather modest overall, and only weakly dependent on
galaxy mass.  These models therefore suggest that supernova blowout
will not be an effective mechanism for eliminating baryons from low
mass galaxies.  

The right panel of Figure~\ref{modelfig} shows that the fraction of
metals lost from the galaxies is a stronger function of galaxy mass,
rising steadily from 10-25\% at $V_c\!=\!200\kms$, to 50-60\% at
$\sim20-30\kms$, with little dependence on the entrainment fraction.
Thus, the fraction of metals varies by less than a factor of 5 over
the entire disk galaxy population, and shows no sharp features at any
particular disk mass.

Even with the larger fraction of metals lost from low mass galaxies,
it is extremely unlikely that dwarf galaxy winds are responsible for
enriching the intergalactic medium (IGM).  Weighting the curves in
Figure~\ref{modelfig} by the baryonic mass function of galaxies
indicates that the dominant source of metals and gas into the IGM
comes from the most massive galaxies, not the dwarfs.  Although low
mass galaxies lose a larger fraction of their metals, the difference
in gas loss is not sufficiently large to make up for their lower
masses overall.  Such galaxies simply do not contribute enough to the
mass budget of galaxies as a whole to have a significant impact on
enrichment of the IGM \citep[see also][]{calura06}. 

Although Figure~\ref{modelfig} confirms that there is some, albeit
weak, dependence of outflow strength on galaxy mass, it remains
premature to conclude that the physical mechanism driving the
correlation is simply the depth of the potential well.  There are many
galaxy properties that correlate with galaxy mass, and some of these
may be more fundamental in limiting the amount of metal-loss.  The
pressure and scale height of the cool ISM and the maximum sizes of HII
regions are known to vary with galaxy mass, and all may play some role
in determining the structure and efficiency of an outflow.  It is also
not yet clear if the metals lost from the cool phase are truly lost
from the galaxy, rather than kept in a bound hot phase.

Finally, the weak dependence of gas loss on galaxy mass in
Figure~\ref{modelfig} is indeed a direct result of the of the
correlation between gas mass fraction and rotation speed, as suggested
in \S\ref{thresholdsec}.  If one ignores this dependence and instead
assumes a constant gas mass fraction characteristic of spirals, then
galaxies with $V_c\!\sim\!90\kms$ would have needed to lose nearly twice
as much gas as calculated above, and galaxies with $V_c\!\sim\!20\kms$
would have had to lose more than 90\% of their gas.  When the
variation of gas mass fraction with rotation speed is included, the
required mass-loss is significantly smaller.

\section{Conclusions}   \label{conclusionsec}

This paper has presented a series of calculations showing the impact
that inflow (\S\ref{infallsec}), outflow (\S\ref{outflowsec}), and
subsequent closed-box evolution (\S\ref{evolutionsec}) have on the
effective yield and gas mass fraction of galaxies.  These
calculations, and their comparison with observations
(\S\ref{pilyuginsec}-\S\ref{modelsec}), yield the following
conclusions:

\begin{enumerate}

\item There is a minimum effective yield that can be produced by gas
   accretion.  The minimum is significantly higher than the effective
   yields observed in low mass galaxies.  Thus low effective yields
   cannot be due to gas infall.

\item Outflows that drive out gas with the mean metallicity of the ISM
   can only produce low effective yields if they remove nearly the
   entire ISM.  Because the same galaxies that show low effective
   yields also show high gas fractions, outflows of unenriched gas
   cannot produce low effective yields.

\item Outflows that consist primarily of escaped SN ejecta are extremely
   efficient at reducing the effective yield.  Metal-enriched outflow
   is therefore the only viable mechanism for producing galaxies with
   low effective yields.

 \item Metal-enriched outflows are ineffective at reducing the
   effective yields of gas-poor systems.  Thus, only galaxies that
   have maintained high gas mass fractions will show depressed
   effective yields in response to outflow.  Only galaxies with
   inefficient star formation are capable of remaining gas-rich to the
   present day, and thus they are the only systems which can show
   significantly depressed effective yields.

\item Any star formation that takes place after gas accretion or gas loss
   will increase the effective yield back to the true nucleosynthetic
   yield expected for closed-box evolution.  Thus, only galaxies with
   inefficient star formation can maintain low effective yields long
   after infall or outflow has finished.

 \item Points 4~\&~5 suggest that in addition to the ability to drive
   winds, galaxies with low effective yields must also have high gas mass
   fractions and low star formation rates.  Thus, the drop in
   effective yield observed at $V_c\!\sim\!120\kms$ ($M_{baryon}\!\sim\!
   10^{10}\msun$) cannot be interpreted solely as a mass threshold for
   escape of SN-driven winds.  More massive galaxies may have also
   experienced significant outflows, but will show no reduction in
   their effective yields due to their low gas mass fractions and
   efficient star formation.

 \item Points 1~\&~2 should not be taken as evidence that infall and
   unenriched outflows have not occurred.  They only indicate that
   these two processes produce little noticeable change in a galaxy's
   effective yield.

 \item At high redshifts, a larger fraction of galaxies are likely to
   be gas-rich.  Therefore a larger fraction of high redshift galaxies
   will show reduced effective yields, even if the rates and sizes of
   typical outflows are identical to those at the present day.

 \item The current data on the effective yield and gas mass fraction
   as a function of galaxy rotation speed suggest that the fraction of
   metals lost due to outflows increases steadily toward lower mass
   galaxies, reaching 50\% at $\lesssim30\kms$.  However, the fraction
   of baryonic mass lost is quite modest ($\lesssim 15$\%) at all
   galaxy masses.  Neither the fraction of metals lost or the fraction
   of baryons lost shows a significant feature at $V_c\!\sim\!120\kms$.

 \item If either the initial mass function or the nucleosynthetic
   yield depends strongly on metallicity, then the amount of gas and
   metal-loss needed to explain the effective yield data would be
   different than calculated above.

\end{enumerate}

\acknowledgements The author gratefully acknowledges discussions with
Andrew West, Alyson Brooks, Anil Seth, Crystal Martin, Yong-Zhong
Qian, Don Garnett, \& Peter Yoachim.  A referee also made several
suggestions which significantly extended and improved the manuscript.

JJD was partially supported through NSF grant CAREER AST-0238683, and
the Alfred P.\ Sloan Foundation. 

\begin{appendix}

\section{Continuous versus Impulsive Chemical Evolution} \label{proofsec}

The calculations in this paper consider the impact of single ``impulsive''
episodes of chemical evolution on the effective yield.  However, gas
accretion, outflow, and star formation are likely to be continuous and
interleaved.  In this Appendix, I show that these impulsive cases are
strict upper and lower bounds on the effective yield produced by more
continuous chemical evolution models.  The proof rests on three lemmas
(proved in \S\ref{lemma1sec}-\ref{lemma3sec}) governing how the
effective yield changes when swapping the order of two sequential impulsive
episodes: (1) star formation followed by gas accretion
always produces a lower effective yield than gas accretion followed by
star formation; (2) star formation followed by gas outflow always
produces a higher effective yield than gas outflow followed by star
formation; (3) sequential episodes of gas accretion and gas outflow
produce identical effective yields, independent of the order of the
two events.  Physically, the first lemma is true because gas
accretion produces the largest drop in the effective yield of gas-poor
systems.  Thus, if star formation occurs first, it reduces the
gas mass fraction and increases the
impact of the infall.  Similarly, the second lemma is true because
outflows produce the largest drop in the effective yield of gas-rich
systems.  Thus, the outflow's impact is largest if it occurs before
star formation has reduced the gas-richness.

With these lemmas, I show that for the accretion, outflow, and
conversion into stars of fixed quantities of gas, the minimum
effective yield is produced when impulsive outflow is followed by
closed-box star formation and then followed by impulsive gas
accretion.  The maximum effective yield is produced when the order of
these processes is reversed (i.e.\ infall, followed by star formation,
followed by outflow).  These extremes allow one to calculate the range
of possible effective yields produced by an arbitrary star formation
and gas flow history.

\subsection{Impulsive Chemical Evolution as a Limiting Case}

Assume that during some interval of time $\Delta M_{stars}$ is
converted from gas into stars, $\Delta M_{infall}$ of gas is accreted
along with $\Delta M_{Z,infall}$ in metals, and $\Delta M_{outflow}$
and $\Delta M_{Z,outflow}$ of gas and metals are carried away by
outflows.  For the most general case, assume that there are no
constraints on the rate and timing of when the gas is accreted or
expelled, or on the rate and timing of when any new stars are formed,
provided that the correct totals are reached at the end of the time
interval.  Given this freedom, how can the timing of the infall,
outflow, and star formation be adjusted to produce the maximum
reduction in the effective yield at the end of the time interval?

An arbitrary continuous chemical evolution history can be approximated
as an interleaved series of infinitesimal impulsive episodes of star
formation, gas infall, and gas outflow, each of which produce changes
$\delta M_{stars,i}$, $\delta M_{infall,i}$, $\delta M_{Z,infall,i}$,
$\delta M_{outflow,i}$ and $\delta M_{Z,outflow,i}$ in the mass of
gas, stars, and metals.  This time sequence can be represented as
\begin{equation}
\mathbf{I_1 O_1 S_1 I_2 O_2 S_2 I_3 O_3 S_3}\cdots
\end{equation}
where $\mathbf{S_i}$ represents impulsive (closed-box) star formation
of $\delta M_{stars,i}$ masses of stars, $\mathbf{I_i}$ represents
impulsive gas accretion of $\delta M_{infall,i}$ masses of gas along
with $\delta M_{Z,infall,i}$ masses of metals, and $\mathbf{O_i}$
represents impulsive outflow of $\delta M_{outflow,i}$ masses of gas
and $\delta M_{Z,outflow,i}$ masses of metals.  These events can be
made arbitrarily small so that the sequence closely approximates a
continuous chemical evolution history.  However, they must obey mass
conservation such that $\Delta M_{infall}\!=\!\sum_i \delta M_{infall,i}$,
$\Delta M_{outflow}\!=\!\sum_i \delta M_{outflow,i}$, and $\Delta
M_{stars}\!=\!\sum_i \delta M_{stars,i}$.

Now consider reordering pairs in the series above.  If some pair is
reordered such that it produces a lower effective yield at that point
in time, then all subsequent effective yields in the series will be
lowered as well, since the final effective yield changes linearly with
the initial effective yield (Equations~\ref{ratiofgeqn},
\ref{ratiofgoutfloweqn},~\&~\ref{yeffpostfloweqn2}).  Thus, given some
arbitrary star formation and accretion history, one can arrange a
lower final effective yield by reordering any adjacent pair of $\delta
M_{infall}$ and $\delta M_{stars}$ so that the gas addition occurs
{\emph{after}} the star formation, or any adjacent pair of $\delta
M_{outflow}$ and $\delta M_{stars}$ so that the outflow occurs
{\emph{before}} the star formation.  Reordering pairs of infall and
outflow episodes makes no change in the final effective yield.

Given the lemmas described above (which are proved below), one can
make the following re-orderings in the series above, switching the
terms in brackets in each step:
\begin{eqnarray*}
&  \underbrace{\mathbf{I_1} \mathbf{O_1}} \mathbf{S_1} \underbrace{\mathbf{I_2} \mathbf{O_2}} \mathbf{S_2} \underbrace{\mathbf{I_3} \mathbf{O_3}} \mathbf{S_3}  & \qquad y_{eff,1} \\
&  \mathbf{O_1} \underbrace{\mathbf{I_1} \mathbf{S_1}} \mathbf{O_2} \underbrace{\mathbf{I_2} \mathbf{S_2}} \mathbf{O_3} \underbrace{\mathbf{I_3} \mathbf{S_3}}  & \qquad y_{eff,2}=y_{eff,1} \\
&  \mathbf{O_1} \mathbf{S_1} \underbrace{\mathbf{I_1} \mathbf{O_2}} \mathbf{S_2} \underbrace{\mathbf{I_2} \mathbf{O_3}} \mathbf{S_3} \mathbf{I_3} &  \qquad y_{eff,3}<y_{eff,2} \\
&  \mathbf{O_1} \underbrace{\mathbf{S_1} \mathbf{O_2}} \underbrace{\mathbf{I_1} \mathbf{S_2}} \mathbf{O_3} \underbrace{\mathbf{I_2} \mathbf{S_3}} \mathbf{I_3}  & \qquad y_{eff,4}=y_{eff,3} \\
&  \mathbf{O_1} \mathbf{O_2} \mathbf{S_1} \mathbf{S_2} \underbrace{\mathbf{I_1} \mathbf{O_3}} \mathbf{S_3} \mathbf{I_2} \mathbf{I_3} &  \qquad y_{eff,5}<y_{eff,4} \\
&  \mathbf{O_1} \mathbf{O_2} \underbrace{\mathbf{S_1} \mathbf{S_2} \mathbf{O_3}} \underbrace{\mathbf{I_1} \mathbf{S_3}} \mathbf{I_2} \mathbf{I_3} &  \qquad y_{eff,6}=y_{eff,5} \\
& \mathbf{O_1} \mathbf{O_2} \mathbf{O_3} \mathbf{S_1} \mathbf{S_2} \mathbf{S_3} \mathbf{I_1} \mathbf{I_2} \mathbf{I_3} & \qquad y_{eff,7}<y_{eff,6}
\end{eqnarray*}
Only in the bottom configuration are there no possible re-orderings
that could produce a lower effective yield.  This final sequence of
outflow, followed by closed-box star formation, followed by infall
therefore produces the minimum effective yield of any chemical
evolution history that involves the same total gas masses as the first
sequence.  The above arguments also prove that the minimum effective
yield that can be produced by infall of a fixed amount of gas is
observed immediately after the infall ceases.

In the same vein, the {\emph{maximum}} effective yield that can be
reached for a given change in the gas and stellar mass can be found by
reordering the initial sequence such that each swap produces a
{\emph{greater}} effective yield.  The end point of that process is
$\mathbf{I_1 I_2 I_3 S_1 S_2 S_3 O_1 O_2 O_3}$.  This result also
prove that the final effective yield for an arbitrary outflow history
is always smaller than if calculated for a single impulsive outflow event.

Taken together, the results above prove that the effective yield
produced by an arbitrary continuous chemical evolution history must
lie between the two extremes of the impulsive outflow-SF-inflow case
and the impulsive inflow-SF-outflow case. 

The only possible complications in the above proof are if either
$\Delta M_{outflow}+\Delta M_{stars} > M_{gas,initial}$ or if $\Delta
M_Z > M_{Z,initial}$ for the minimum effective yield case.  If so,
there would not be enough gas and/or metals in the initial gas
reservoir to support the outflow and subsequent star formation.  In
these cases, the process of reordering to reach a lower effective
yield would be limited by the need to maintain a positive gas mass at
all times.  However, the same principles apply, and an optimal
reordering could be reached for the specific case under consideration.

\subsection{Infall versus Closed-Box Star Formation}  \label{lemma1sec}

To prove that the effective yield is lower if $\delta M_{stars}$ of
gas converts into stars before $\delta M_{gas}$ is accreted, consider
two cases.  In Case~1, the gas accretion occurs first, and is then
followed by closed-box star formation.  In Case~2, the same events
occur in the opposite order.  Assume that the initial gas mass
fraction and effective yield are $f_{gas}$ and $y_{eff}$,
respectively, and that the final gas mass fraction is $f_{gas,f}$.
Assume that the intermediate gas mass fractions (i.e. between the star
formation and gas accretion episodes) are $f_{gas,1}$ for Case~1 and
$f_{gas,2}$ for Case~2.  In terms of the initial gas mass fraction,
\begin{equation}
f_{gas,1} = f_{gas} \frac{1+Y}{1+Yf_{gas}}
\hskip 1cm
f_{gas,2} = f_{gas} (1-XY)
\hskip 1cm
f_{gas,f} = f_{gas,1}\frac{1+Y-XY}{1+Y}
\end{equation}
where $Y\equiv\delta M_{gas}/M_{gas,initial}$ (i.e.\ the fractional
change in the gas mass; see Equation~\ref{fgaseqn}) and $X\equiv\delta
M_{stars}/\delta M_{gas}$.  Also assume that $\Delta Z\equiv \delta
M_Z/M_Z$, where $M_Z$ is the initial mass in metals, and $\delta M_Z$
is the mass of accreted metals.

For Case~1, the intermediate gas mass fraction $f_{gas,1}$ can be used
in Equation~\ref{ratiofgeqn} to derive the intermediate effective
yield after the gas accretion.  The intermediate effective yield and
gas mass fraction can then be used to calculate the final effective yield
$y_{eff,1}$ after the subsequent star formation, using a modified
form of Equation~\ref{yeffpostfloweqn3}:
\begin{equation}                     \label{modifiedSFeqn}
\frac{y_{eff}}{y_{eff,postflow}} = 
     \left[\frac{\ln{f_{gas,postflow}}}{\ln{f_{gas}}}\right]
     \left[\left(\frac{\ln{f_{gas}}}{\ln{f_{gas,postflow}}}-1\right)
	        \left(\frac{y_{true}}{y_{eff,postflow}}\right) - 1\right].
\end{equation}
Setting $y_{eff,postflow}$ and $f_{gas,postflow}$ to the values after
gas accretion, the final effective yield $y_{eff,1}$ for Case~1 is
\begin{eqnarray}
\frac{y_{eff,1}}{y_{eff}} &=&
          \left[1+\Delta Z\right]
          \left[\frac{\ln{f_{gas}}}{\ln{f_{gas,f}}}\right]
          \left[\frac{f_{gas}}{f_{gas,1}}\right]
          \left[\frac{1-f_{gas,1}}{1-f_{gas}}\right] \times \nonumber \\
          & & \left[\left(\frac{\ln{f_{gas,f}}}{\ln{f_{gas,1}}}-1\right)
                \left(\frac{\ln{f_{gas,1}}}{\ln{f_{gas}}}\right)
                \left(\frac{f_{gas,1}}{f_{gas}}\right)
                \left(\frac{1-f_{gas}}{1-f_{gas,1}}\right)
                \left(\frac{1}{1+\Delta Z}\right)
	        \left(\frac{y_{true}}{y_{eff}}\right) - 1 \right].
\end{eqnarray}

For Case~2, Equation~\ref{ratiofgeqn} and the modified form of
Equation~\ref{yeffpostfloweqn3} can be applied
in the opposite order as in Case~1.  The final effective yield
$y_{eff,2}$ for Case~2 is then
\begin{equation}
\frac{y_{eff,2}}{y_{eff}}=
          \left[1+\Delta Z\left(\frac{M_Z}{M_{Z,2}}\right)\right]
          \left[\frac{\ln{f_{gas}}}{\ln{f_{gas,f}}}\right]
          \left[\frac{f_{gas,2}}{f_{gas,f}}\right]
          \left[\frac{1-f_{gas,f}}{1-f_{gas,2}}\right]
          \left[\left(\frac{\ln{f_{gas,2}}}{\ln{f_{gas}}}-1\right)
	        \left(\frac{y_{true}}{y_{eff}}\right) - 1\right],
\end{equation}
where $M_{Z,2}$ is the mass in metals after the initial episode of
star formation.

With the above relations, one can compare the final effective yield
when the gas accretion occurs first (Case~1) to the final effective
yield when the gas accretion occurs last (Case~2).  Taking the ratio
of the final effective yields in the two cases, and using the
simplifying expressions
\begin{equation}
\frac{\ln{f_{gas,f}}}{\ln{f_{gas,1}}} - 1 = 
       \frac{\ln{\left(1-\frac{XY}{1+Y}\right)}}{\ln{f_{gas,1}}}
\qquad
\frac{\ln{f_{gas,2}}}{\ln{f_{gas}}} - 1 = 
       \frac{\ln{\left(1-XY\right)}}{\ln{f_{gas}}}
\end{equation}
the ratio of the final effective yields of the two cases becomes
\begin{eqnarray}
\frac{y_{eff,1}}{y_{eff,2}} &=& 
     \left[\frac{1+\Delta Z}{1+\Delta Z\left(\frac{M_Z}{M_{Z,2}}\right)}\right]
     \left[\frac{f_{gas}f_{gas,f}}{f_{gas,1}f_{gas,2}}\right]
     \left[\frac{(1-f_{gas,1})(1-f_{gas,2})}{(1-f_{gas})(1-f_{gas,f})}\right]
     \times \nonumber \\
     & & \left[\frac{1+\frac{A}{\ln{f_{gas}}}
                   \ln{\left(1-\frac{XY}{1+Y}\right)}
                  \left(\frac{1}{1+\Delta Z}\right)
                   \left(\frac{1-f_{gas}}{f_{gas}}\right)
                   \left(\frac{f_{gas,1}}{1-f_{gas,1}}\right)}
                {1 + \frac{A}{\ln{f_{gas}}}
                   \ln{\left(1-XY\right)}}\right]
\end{eqnarray}
where $A\equiv y_{true}/y_{eff}$.  The third term in brackets on the
right hand side is a product of several stellar mass fractions, and is
equal to 1.  Within the fourth term in brackets,
$\left(\frac{1-f_{gas}}{f_{gas}}\right)
\left(\frac{f_{gas,1}}{1-f_{gas,1}}\right)$ simplifies to $1+Y$,
reducing the ratio of $y_{eff,1}/y_{eff,2}$ to
\begin{equation}                         \label{caseratioeqn}
\frac{y_{eff,1}}{y_{eff,2}} = 
     \left[\frac{1+\Delta Z}{1+\Delta Z\left(\frac{M_Z}{M_{Z,2}}\right)}\right]
     \left[\frac{f_{gas}f_{gas,f}}{f_{gas,1}f_{gas,2}}\right]
     \left[\frac{1 + \frac{A}{\ln{f_{gas}}}
                  \left(\frac{1+Y}{1+\Delta Z}\right)
                   \ln{\left(1-\frac{XY}{1+Y}\right)}}
          {1+\frac{A}{\ln{f_{gas}}}
                   \ln{\left(1-XY\right)}}\right].
\end{equation}

If Case~2 indeed produces a lower effective yield than Case~1, the
ratio of $y_{eff,1}/y_{eff,2}$ should always be greater than 1.  The
first term in brackets on the right hand side of
Equation~\ref{caseratioeqn} is always greater than 1,
since the metal mass after star formation $M_{Z,2}$ must be greater than
the initial mass in metals $M_Z$.  The
second term in brackets is equal to $(1+Y-XY)/(1+Y)(1-XY) =
(1+Y-XY)/(1+Y-XY-XY^2)$, which is always greater than or equal to 1,
due to the additional $-XY^2$ term in the denominator.  The third
term is always greater than 1 as well, which can be seen by expanding
the logarithmic terms as a series:
\begin{eqnarray}     
\frac{y_{eff,1}}{y_{eff,2}} &=& 
     \left[\frac{1+\Delta Z}{1+\Delta Z\left(\frac{M_Z}{M_{Z,2}}\right)}\right]
     \left[\frac{f_{gas}f_{gas,f}}{f_{gas,1}f_{gas,2}}\right]
     \times \nonumber \\
     & &\left[\frac{1 - \frac{AXY}{\ln{1/f_{gas}}}
                  \left(\frac{1}{1+\Delta Z}\right)
                   \left(1 + \frac{1}{2}\frac{XY}{1+Y} 
		           + \frac{1}{3}\frac{(XY)^2}{(1+Y)^2} 
			   + \cdots \right)}
          {1 - \frac{AXY}{\ln{1/f_{gas}}}
                   \left(1 + \frac{1}{2} XY 
		           + \frac{1}{3}(XY)^2 + \cdots \right)}\right]
\end{eqnarray}
Since $1+Y$ is always greater than 1, each term in the series
expansion in the numerator is less than or equal to the corresponding
term in the expansion in the denominator.  Because the prefactor
$AXY/\ln{1/f_{gas}}$ is always positive and $(1+\Delta Z)^{-1}<1$, the
numerator of the second term must therefore be greater than the
denominator, for any value of the the initial effective yield and gas
fraction.  All three of the bracketed terms on the right hand side
are therefore greater than 1.  The product of the three terms must
also be greater than 1 as well, thus proving that
$y_{eff,1}/y_{eff,2}>1$ in all cases, as required to prove that the
maximum drop in the effective yield is produced when a system is
observed immediately after gas accretion.  Physically, accretion has
the largest impact on the effective yield when a system is gas-poor.
Thus, any star formation that takes place before the infall would
reduce the gas richness, making the infall more effective in reducing
the effective yield.

\subsection{Outflow versus Closed-Box Star Formation}   \label{lemma2sec}

To prove that the effective yield is higher if $\delta M_{stars}$ of
gas converts into stars before $\delta M_{gas}$ and $\delta M_{Z}$
in gas and metals are
expelled, consider two cases.  In Case~1, the outflow occurs
first, and is then followed by closed-box star formation.  In Case~2,
the same events occur in the opposite order.  Adopting the same notation
as the previous section, and assuming
that the intermediate gas mass fractions (i.e. between the star
formation and outflow episodes) are $f_{gas,1}$ for Case~1 and
$f_{gas,2}$ for Case~2,
\begin{equation}
f_{gas,1} = f_{gas} \frac{1-Y}{1-Yf_{gas}}
\hskip 1cm
f_{gas,2} = f_{gas} (1-XY)
\hskip 1cm
f_{gas,f} = f_{gas,1}\frac{1-Y-XY}{1-Y}
\end{equation}
where $Y\equiv\delta M_{gas}/M_{gas,initial}$ (i.e.\ the fractional
change in the gas mass) and $X\equiv\delta M_{stars}/\delta M_{gas}$.
Also assume that $\Delta Z\equiv \delta M_Z/M_Z$, where $M_Z$ is the
initial mass in metals, and $\delta M_Z$ is the mass lost in metals
during the outflow.  With these definitions, and applying
Equation~\ref{ratiofgoutfloweqn} (with $\Delta Z$ substituted for $x
\Delta M_g/M_g$) and Equation~\ref{modifiedSFeqn},
\begin{eqnarray}
\frac{y_{eff,1}}{y_{eff}} &=&
          \left[\frac{\ln{f_{gas}}}{\ln{f_{gas,f}}}\right]
          \left[\frac{1-\Delta Z}{1-Y}\right] \times \nonumber \\
          & & \left[\left(\frac{\ln{f_{gas,f}}}{\ln{f_{gas,1}}}-1\right)
                \left(\frac{\ln{f_{gas,1}}}{\ln{f_{gas}}}\right)
                \left(\frac{1-Y}{1-\Delta Z}\right)
	        \left(\frac{y_{true}}{y_{eff}}\right) - 1 \right].
\end{eqnarray}

For Case~2, Equation~\ref{ratiofgoutfloweqn} and
Equation~\ref{modifiedSFeqn} can be applied in the opposite order as
in Case~1.  The final effective yield $y_{eff,2}$ for Case~2 is then
\begin{equation}
\frac{y_{eff,2}}{y_{eff}}=
          \left[\frac{\ln{f_{gas}}}{\ln{f_{gas,f}}}\right]
          \left[\frac{1-\Delta Z\left(\frac{M_{Z}}{M_{Z,2}}\right)}
                     {1-Y}\right]
          \left[\left(\frac{\ln{f_{gas,2}}}{\ln{f_{gas}}}-1\right)
	        \left(\frac{y_{true}}{y_{eff}}\right) - 1\right],
\end{equation}
where $M_{Z,2}$ is the mass in metals after star formation, but before
outflow, such that $M_{Z,2}>M_Z$ in all cases.

With the above relations, one can compare the final effective yield
when the outflow occurs first (Case~1) to the final effective
yield when the outflow occurs last (Case~2).  Taking the ratio
of the final effective yields in the two cases, and using the
simplifying expressions
\begin{equation}
\frac{\ln{f_{gas,f}}}{\ln{f_{gas,1}}}-1 = 
       \frac{\ln{\left(1-\frac{XY}{1-Y}\right)}}{\ln{f_{gas,1}}}
\hskip 2cm
\frac{\ln{f_{gas,2}}}{\ln{f_{gas}}}-1 = 
       \frac{\ln{\left(1-XY\right)}}{\ln{f_{gas}}}
\end{equation}
the ratio of the final effective yields of the two cases becomes

\begin{equation}
\frac{y_{eff,1}}{y_{eff,2}} = 
     \left[\frac{1-\Delta Z}
                {1-\Delta Z \left(\frac{M_{Z}}{M_{Z,2}}\right)} \right]
     \left[\frac{1-\frac{A}{\ln{f_{gas}}}
                   \left(\frac{1-Y}{1-\Delta Z}\right)
                   \ln{\left(1-\frac{XY}{1-Y}\right)}}
                {1 - \frac{A}{\ln{f_{gas}}}
                   \ln{\left(1-XY\right)}}\right]
\end{equation}
where $A\equiv y_{true}/y_{eff}$.

The first term in brackets is always less than 1, because $M_Z/M_{Z,2}<1$.
The second term in brackets is always less than 1 as well, which can be
seen by expanding the logarithmic terms in a series:
\begin{equation}     
\frac{y_{eff,1}}{y_{eff,2}} = 
     \left[\frac{1-\Delta Z}
                {1-\Delta Z \left(\frac{M_{Z}}{M_{Z,2}}\right)} \right]
     \left[\frac{1 - \frac{AXY}{\ln{1/f_{gas}}}
                   \left(\frac{1}{1-\Delta Z}\right)
                   \left(1 + \frac{1}{2}\frac{XY}{1-Y} 
		           + \frac{1}{3}\frac{(XY)^2}{(1-Y)^2} 
			   + \cdots \right)}
          {1 - \frac{AXY}{\ln{1/f_{gas}}}
                   \left(1 + \frac{1}{2} XY 
		           + \frac{1}{3}(XY)^2 + \cdots \right)}\right].
\end{equation}
Since $1-Y<1$, $\Delta Z<1$, and $AXY/\ln{(1/f_{gas})}$ is always
positive, the second term is always less than one.  The product of the
two terms is therefore always less than one as well, thus proving that
$y_{eff,1}/y_{eff,2}<1$ in all cases.  Thus, the minimum effective
yield is produced when star formation follows outflow.  Physically,
outflow has the largest impact on the effective yield when a system is
gas rich.  Thus, any star formation that takes place before outflow
would reduce the gas richness, making the outflow much less effective
in reducing the effective yield.

\subsection{Infall versus Outflow}      \label{lemma3sec}

Assume that there are back-to-back accretion and outflow events.
Assume that a mass $\delta M_{infall}$ of gas and $\delta M_{Z,infall}$ of
metals is accreted and that a mass of $\delta M_{outflow}$ of gas and $\delta
M_{Z,outflow}$ of metals is expelled via outflows.  Let Case~1 be
when outflow occurs first, let Case~2 be when infall occurs first, and
assume the same notation as in previous sections.

In Case~1 the effective yield after the initial outflow is
\begin{equation}
y_{eff,1,intermediate} =  \frac{M_Z -\delta M_{Z,outflow}}
                  {M_{gas} - \delta M_{outflow}}
         \left[ \frac{1}
               {\ln{\left(\frac{M_{stars} + M_{gas} - \delta M_{outflow}}
               {M_{gas} - \delta M_{outflow}}\right)}}
              \right].
\end{equation}
After the subsequent infall, the effective yield becomes
\begin{equation}
y_{eff,1} =  \frac{M_Z -\delta M_{Z,outflow}+\delta M_{Z,infall}}
                  {M_{gas} - \delta M_{outflow} + \delta M_{infall}}
         \left[ \frac{1}
               {\ln{\left(
            \frac{M_{stars} + M_{gas} - \delta M_{outflow} + \delta M_{infall}}
                 {M_{gas} - \delta M_{outflow} + \delta M_{infall}}\right)}}
              \right].
\end{equation}
A similar calculation for Case~2 yields a different intermediate
effective yield, but the final effective yield is identical (i.e.\
$y_{eff,2}=y_{eff,1}$).  Thus, the order in which infall and outflow
occur has no impact on the final effective yield.  This symmetry did
not occur for the two previous cases, because the fraction of metals
locked up by star formation depended on the initial metallicity of the
gas, not just on the total mass of gas converted into stars.

\section{Garnett Data}          \label{garnettsec}

In this section I reproduce the plots from
Figures~\ref{pilyugindatafig},~\ref{pilyuginfig},~\&~\ref{pilyuginevolutionfig}
using the data from the original \citet{garnett02} paper rather than
\citet{pilyugin04}.  \citet{garnett02} used the $R_{23}$ technique to
derive oxygen abundances, and adopted different mass-to-light ratios,
N(H$_2$)/I(CO) conversion factors, and rotation speeds than
\citet{pilyugin04}.  Because of the very different metallicity scales
adopted by these two papers, the data could not be combined on a
single plot in the main body of the paper.  These data are included
for completeness, but do not substantively change the results of the
paper.

Figure~\ref{garnettdatafig} shows the effective yield as a
function of rotation speed (left panel) and gas mass fraction (center
panel) for the data compiled in Table~4 of \citet{garnett02}.  Like
the \citet{pilyugin04} sample, the \cite{garnett02} sample contains
spirals and irregular galaxies drawn from the literature, rectifies
all metallicity measurements to a common abundance calibration, and
interpolates the resulting metallicities to a common galactocentric
radius (the half-light radius for \citet{garnett02} versus $0.4R_{25}$
for \citet{pilyugin04}).

\placefigure{garnettdatafig}

The right panel of Figure~\ref{pilyugindatafig} shows a direct
comparison between the \citet{pilyugin04} and \citet{garnett02}
samples, which have many galaxies in common.  \citet{garnett02}
adopted systematically higher effective yields and higher gas mass
fractions.  The low effective yields are due primarily to the much
higher abundances derived by \citet{garnett02} using the popular
``$R_{23}$'' method \citep{pagel79}.

\citet{garnett02} also make different assumptions when deriving gas
mass fractions.  \citet{pilyugin04} assumed a constant stellar
mass-to-light ratio of $M/L=1.5$ for spirals whereas \citet{garnett02}
uses a more accurate color-dependent mass-to-light ratio.  Both papers
adopt a constant stellar mass-to-light ratio of $M/L=1.0$ for all
irregulars.  \citet{garnett02} also adopts a larger N(H$_2$)/I(CO)
conversion factor, and includes a correction for H$_2$ in irregulars
that was neglected in \citet{pilyugin04}, but that has been included
in Figure~\ref{pilyugindatafig}.  Many of the distances adopted by
both papers vary as well, although this affects only the absolute
magnitude, but not the gas fraction or the effective yield.

\placefigure{garnettfig}

Figure~\ref{garnettdatafig} suggests a higher true nucleosynthetic
yield of $y_{true}=0.01$, rather than
${\rm{log_{10}}}(y_{true})\approx-2.5$ as adopted throughout the
paper.  This higher value is in good agreement with theoretical
calculations by \citet{maeder92} and \citet{nomoto97}, but slightly
higher than the models of \citet{woosley95}, as compiled by
\citet{henry00} for a Salpeter initial mass function.  However, since
the results in this paper calculate the ratio of the final to initial
effective yield, an overall decrease in the metallicity scale has no
effect on our conclusions.  On the other hand, because the metallicity
calibration adopted by \citet{pilyugin04} differs from the $R_{23}$
method primarily at high metallicities, the full range of effective
yields in the \citet{pilyugin04} sample is a factor of two smaller
than the range seen in \citet{garnett02}, reducing the amount of
inflow or outflow needed to explain the \citet{pilyugin04} data.
Figure~\ref{garnettfig} reproduces the flow models previously applied
to Figure~\ref{pilyuginfig} and Figure~\ref{pilyuginevolutionfig}, and
reach the same qualitative conclusions.  With the larger range of
effective yields seen in the \citet{garnett02} sample, metal-enriched
outflows from gas-rich galaxies are even more needed to explain the
data.

\end{appendix}



\ifprintfig

\begin{figure}[p]
\centerline{
\includegraphics[width=2.2in]{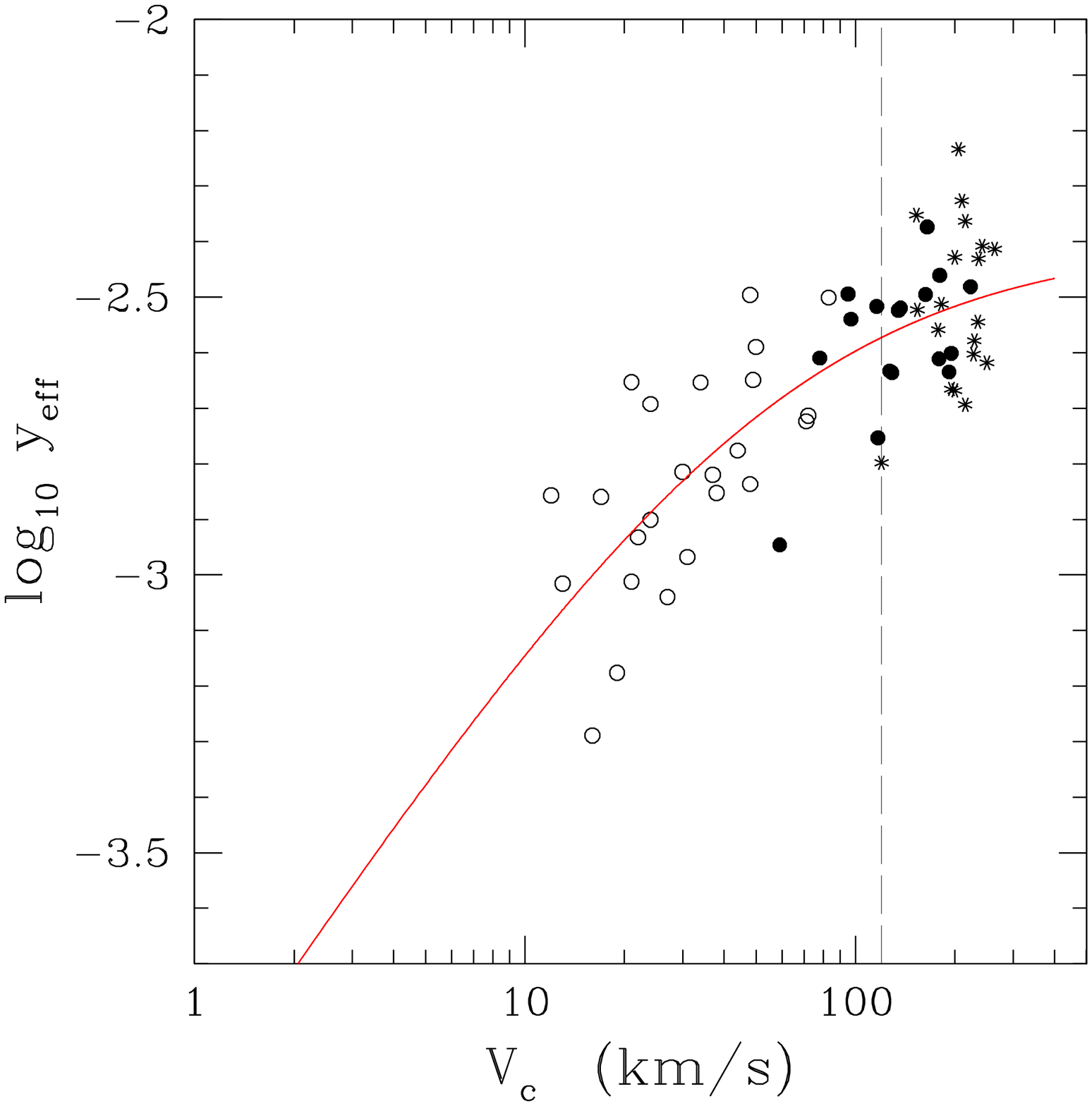}
\includegraphics[width=2.2in]{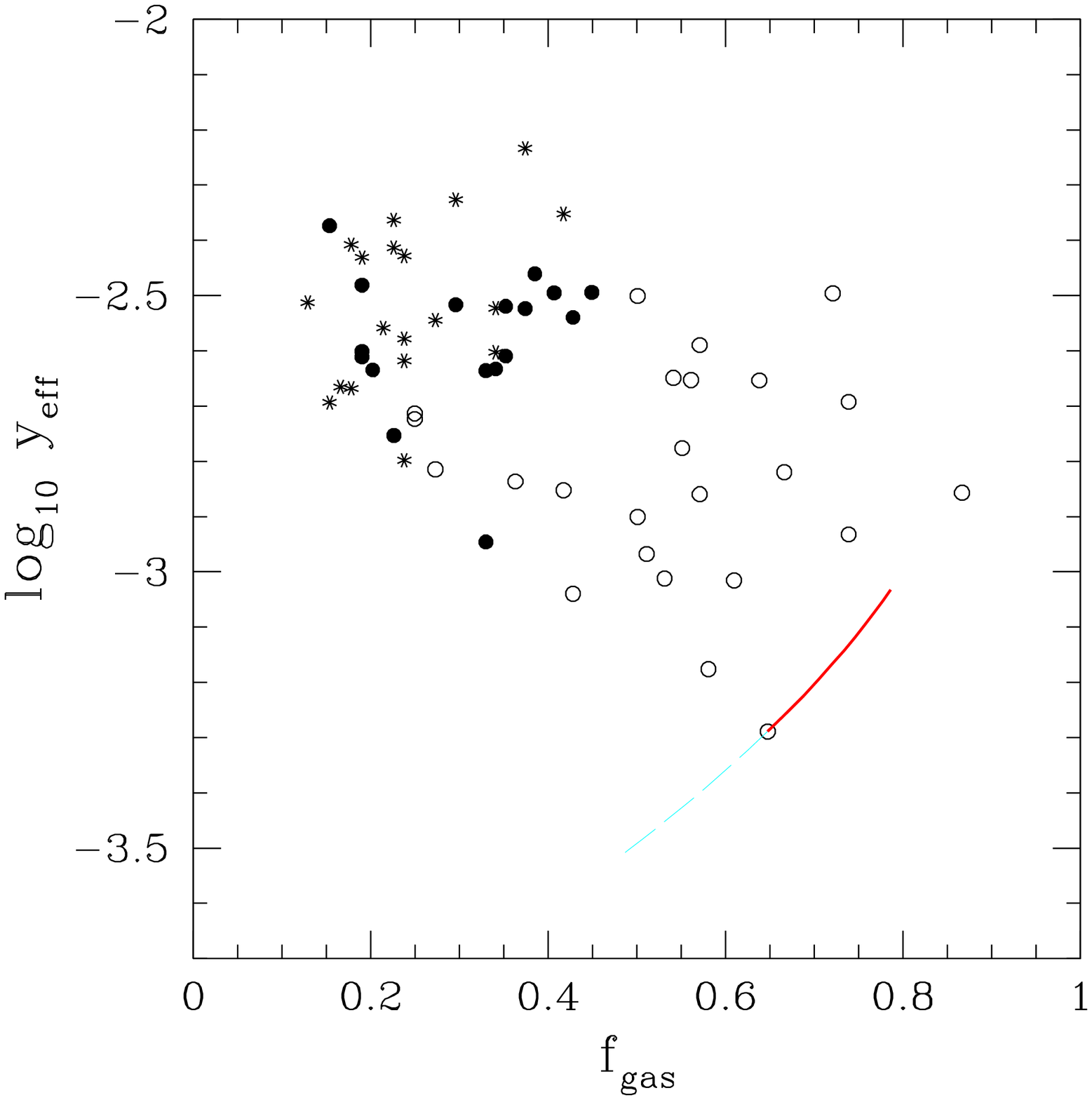}
\includegraphics[width=2.2in]{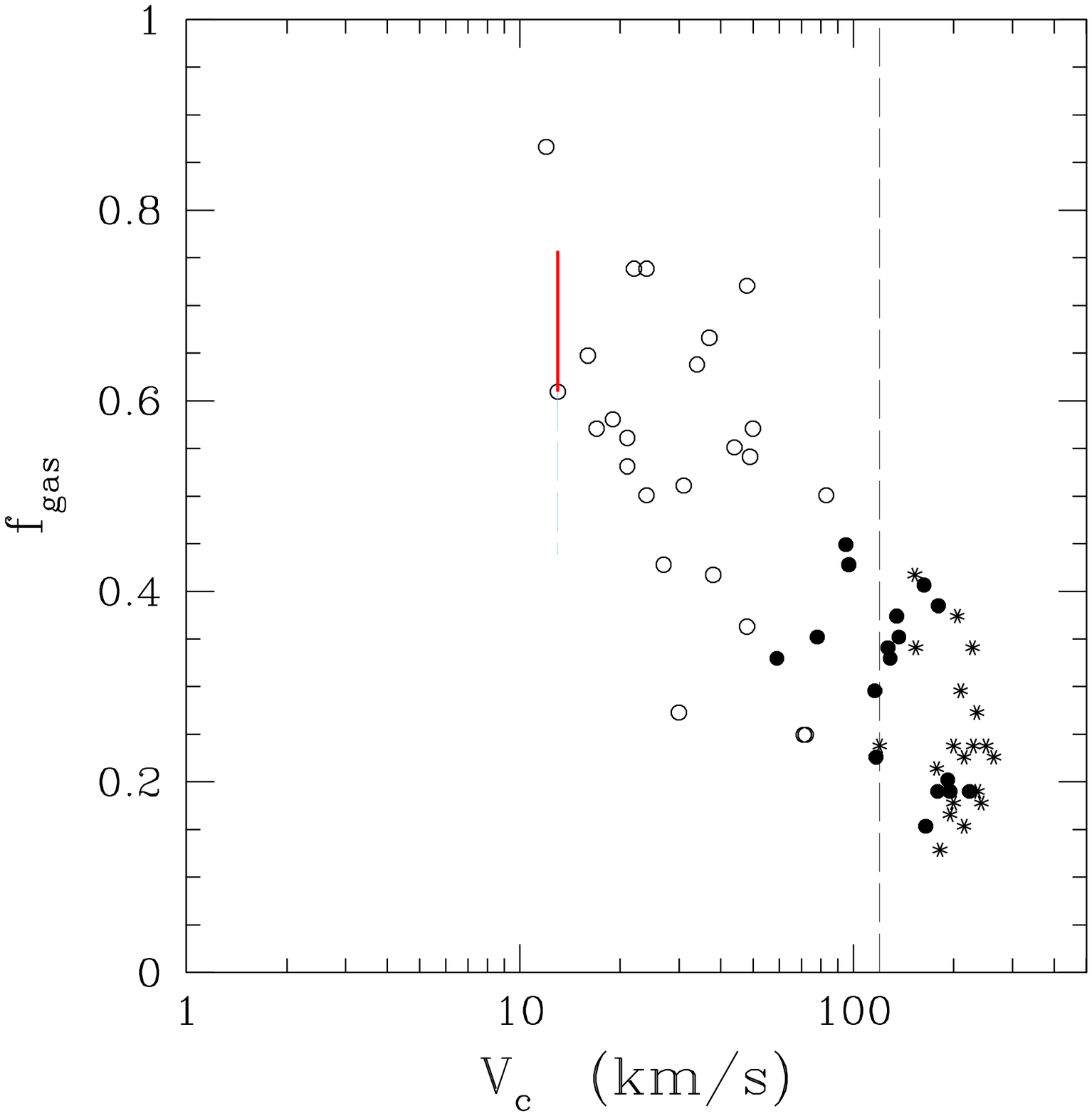}
}
\centerline{
}
\caption{Effective yield as a function of galaxy
         rotation speed (left panel) and of gas mass fraction (center
         panel), measured at $0.4R_{25}$ by \citet{pilyugin04}.  The right
         panel shows the relationship between gas fraction and
         rotation speed for the same data.  Mid-type spirals (Sbc-Sc)
         are plotted as stars, late-type spirals (Scd-Sd) as solid
         circles, and irregulars as open circles.  Compared to massive
         spirals, the effective yield is reduced by nearly a factor
         of ten in low mass galaxies ($V_c<40\kms$), all of which are
         gas-rich ($f_{gas}>0.3$).  In the left and right panels, the
         vertical dashed line indicates the rotation speed below which
         dust lanes disappear and star formation becomes inefficient
         \citep{dalcanton04,verde02}.  In the center and right panels,
         the short lines shows how $y_{eff}$ and $f_{gas}$ would
         change if the measurement of the gas mass (dashed line) or
         the stellar mass (solid line) were reduced by a factor of
         two. The solid curve in the left panel shows the fitting formula
	 adopted in \S\ref{modelsec}.
	 \label{pilyugindatafig}}
\end{figure}
\vfill

\begin{figure}[p]
\centerline{
\includegraphics[width=2.2in]{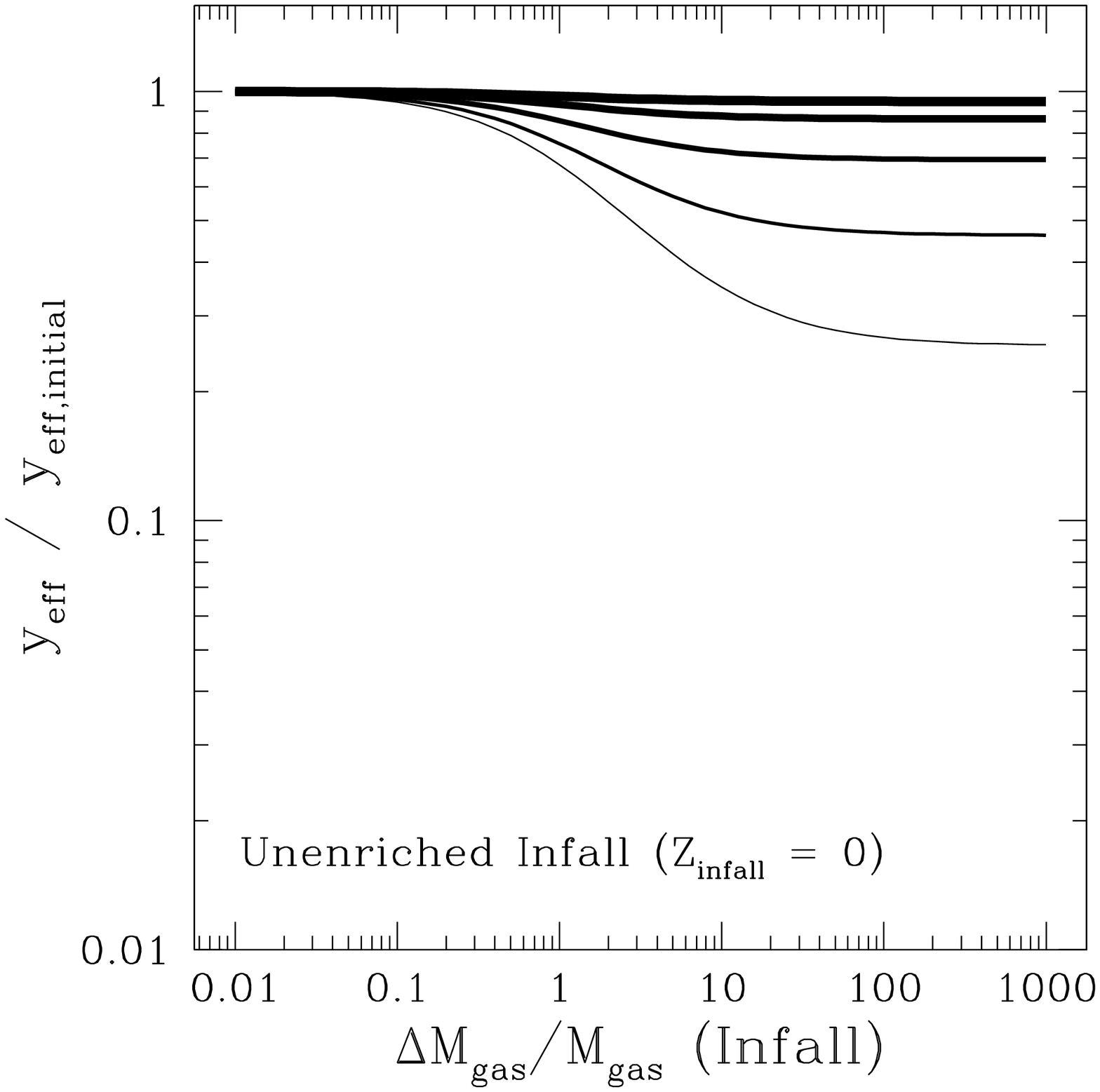}
\includegraphics[width=2.2in]{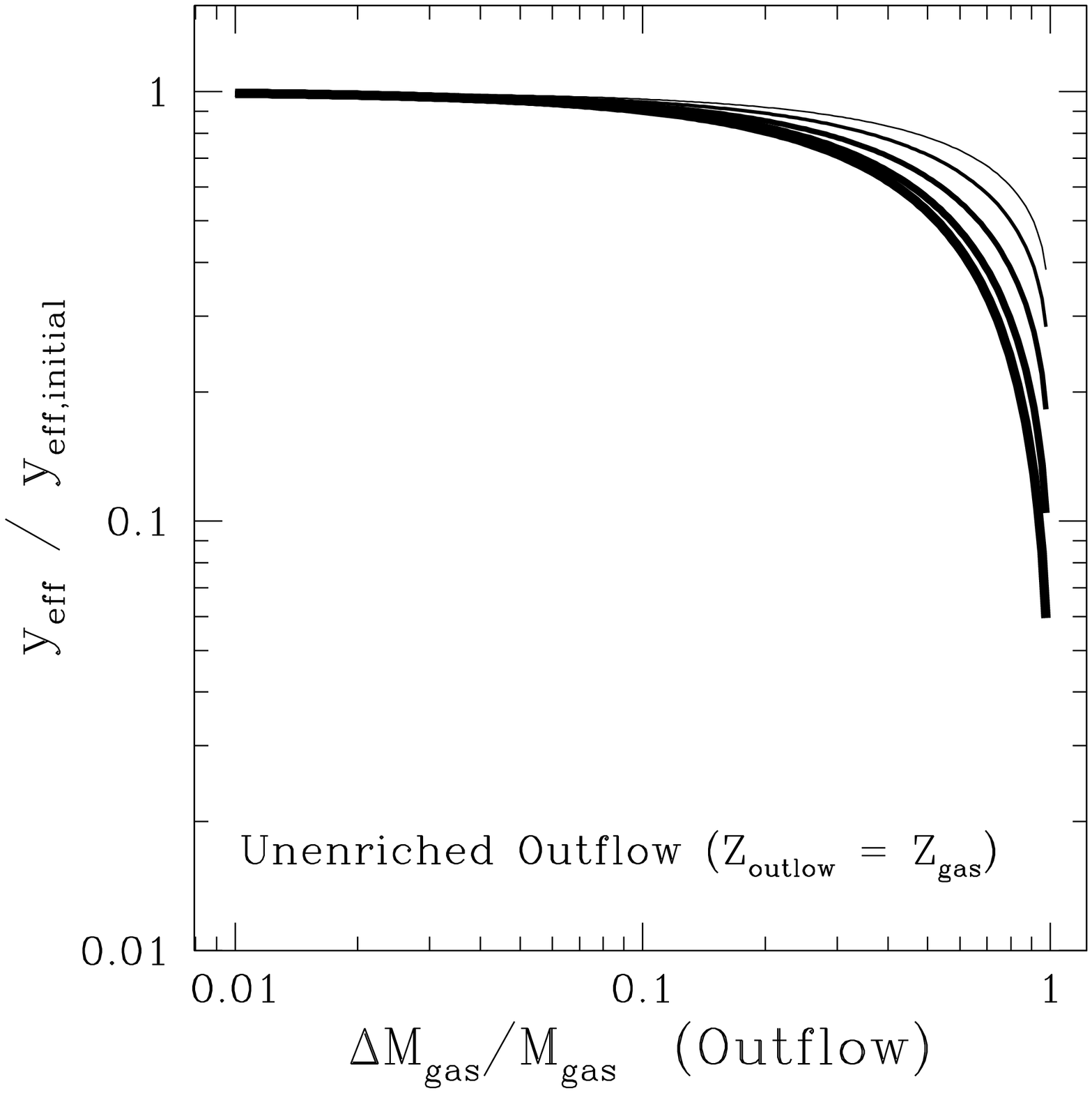}
\includegraphics[width=2.2in]{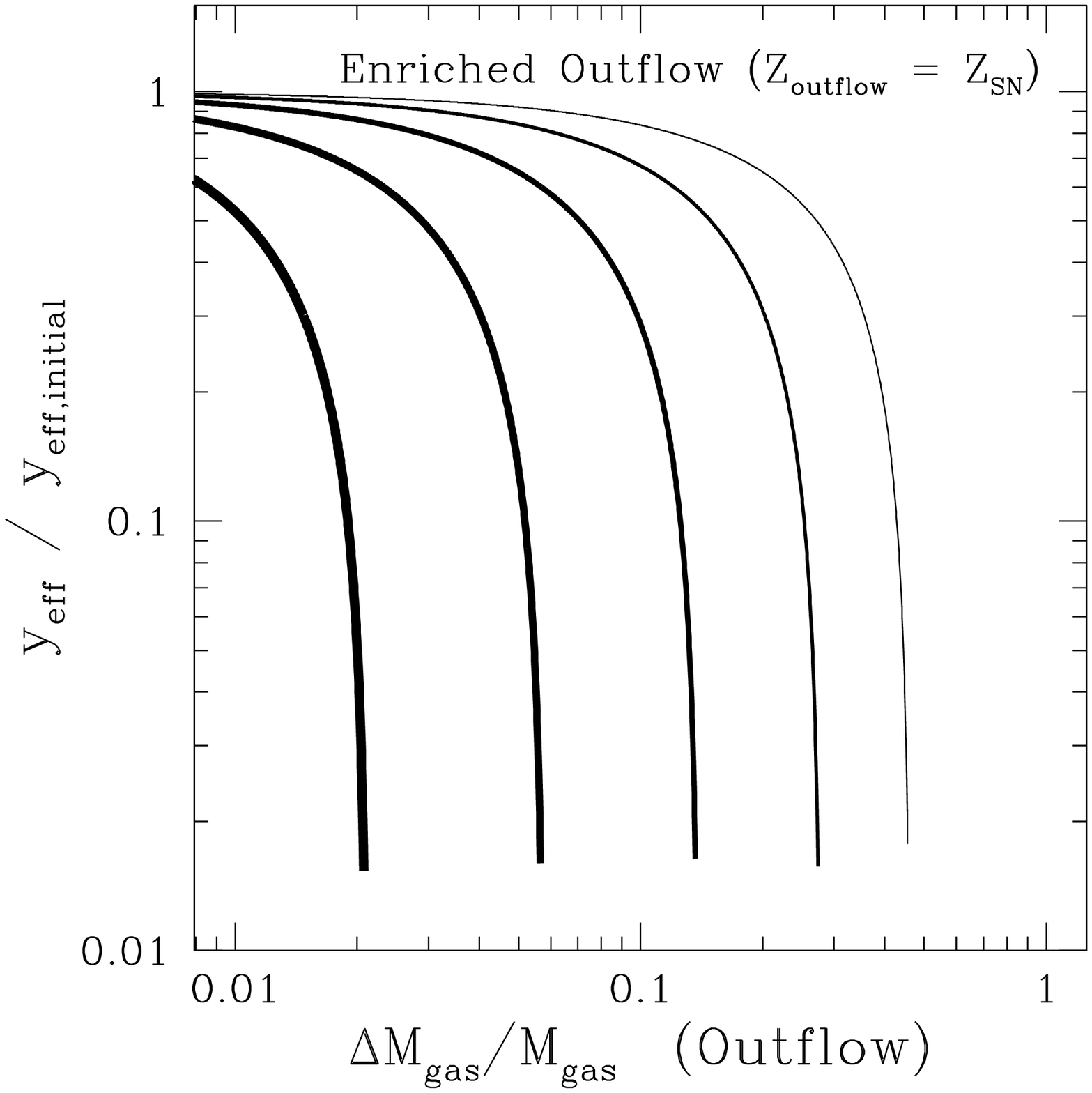}
}
\caption{The ratio of final to initial effective yield as a function
         of the ratio of the fractional change in gas mass $\Delta
         M_{gas}/M_{gas,initial}$, for unenriched infall (left panel;
         $Z_{infall}=0$), unenriched outflow (center panel;
         $Z_{outflow}=Z_{gas}$), and enriched outflow (right panel;
         $Z_{outflow}=Z_{SN}$).  The solid lines show different
         initial gas mass fractions ($f_{gas,initial}=0.1, 0.25, 0.5,
         0.75, 0.9$, plotted from thin to thick, respectively, which
         is from bottom to top in the left panel, and from top to
         bottom in the right panels).  With infall, the effective
         yield is suppressed when large amounts of gas are accreted
         onto galaxies with low initial gas mass fractions.  In the
         two outflow cases, the effective yield is reduced by large
         outflows from galaxies with higher initial gas mass
         fractions.  For unenriched outflows, a factor of ten decrease
         in $y_{eff}$ can only be achieved with nearly total gas
         ablation from initially gas-rich systems.  For enriched
         outflows with large initial gas mass fractions, only
         modest amounts of enriched outflow are required to produce
         the factor of ten drop in the effective yield seen in
         Figure~\ref{pilyugindatafig}.
	 \label{yieldratfig}}
\end{figure}
\vfill
\clearpage

\begin{figure}[p]
\centerline{
\includegraphics[width=6.5in]{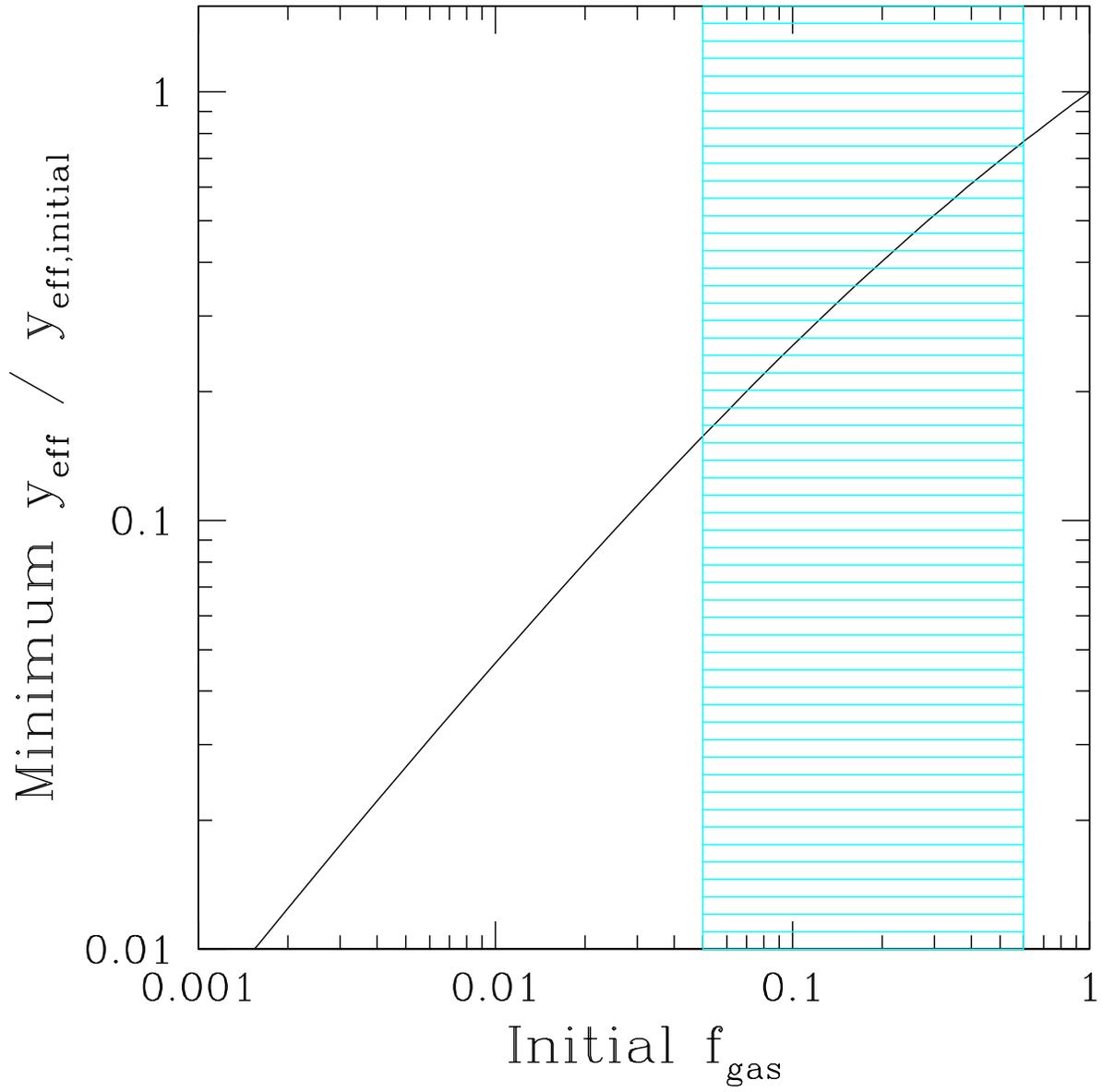}
}
\caption{The minimum possible effective yield that can be produced by
         gas accretion, as a function of the initial gas mass of the
         system.  The shaded region indicates the range of gas
         fractions seen in HI-selected galaxies from the HIPASS survey
         \citep{west05}.
  \label{ratiominimumfig}}
\end{figure}
\vfill
\clearpage

\begin{figure}[p]
\centerline{
\includegraphics[width=2.2in]{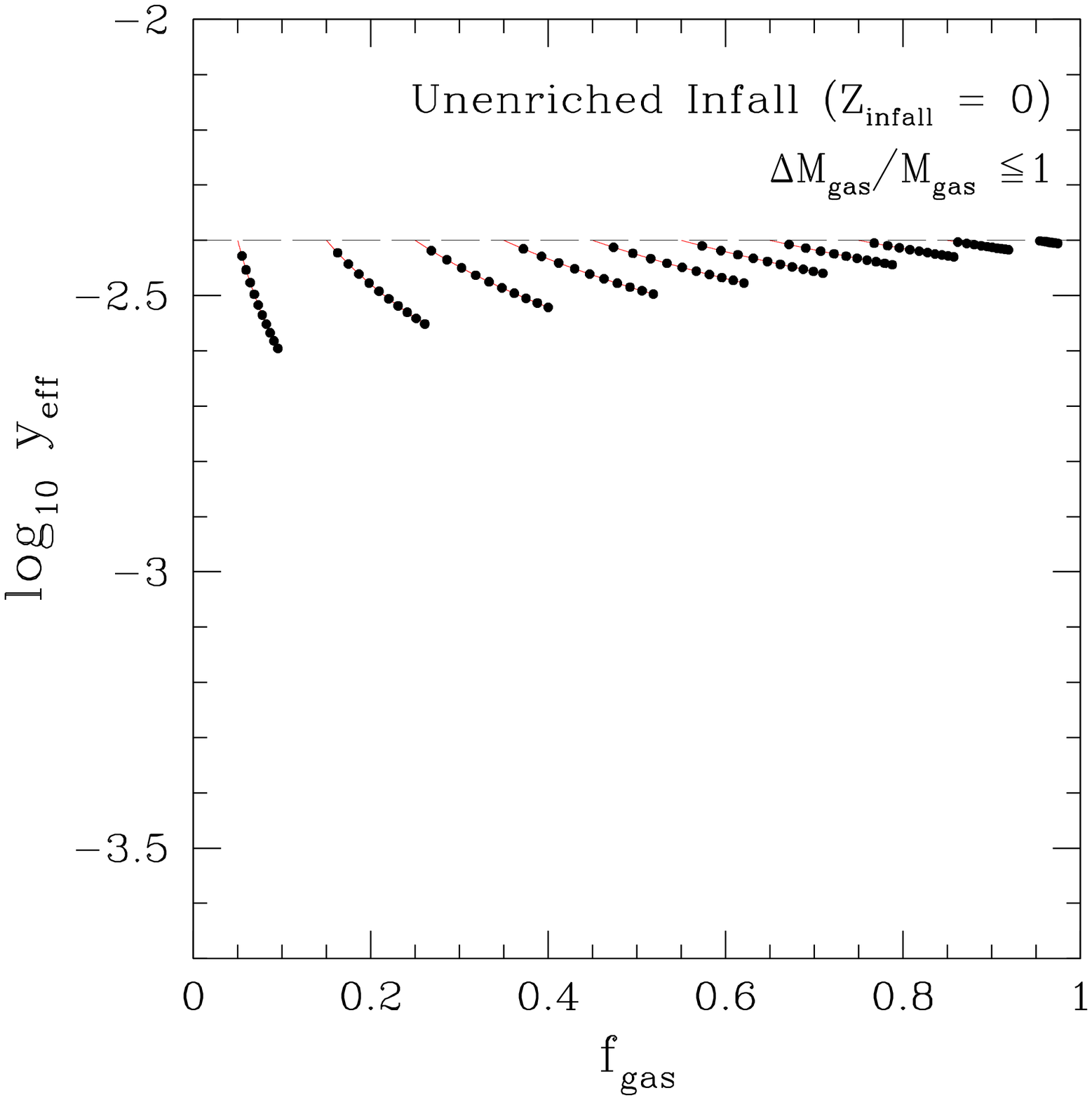}
\includegraphics[width=2.2in]{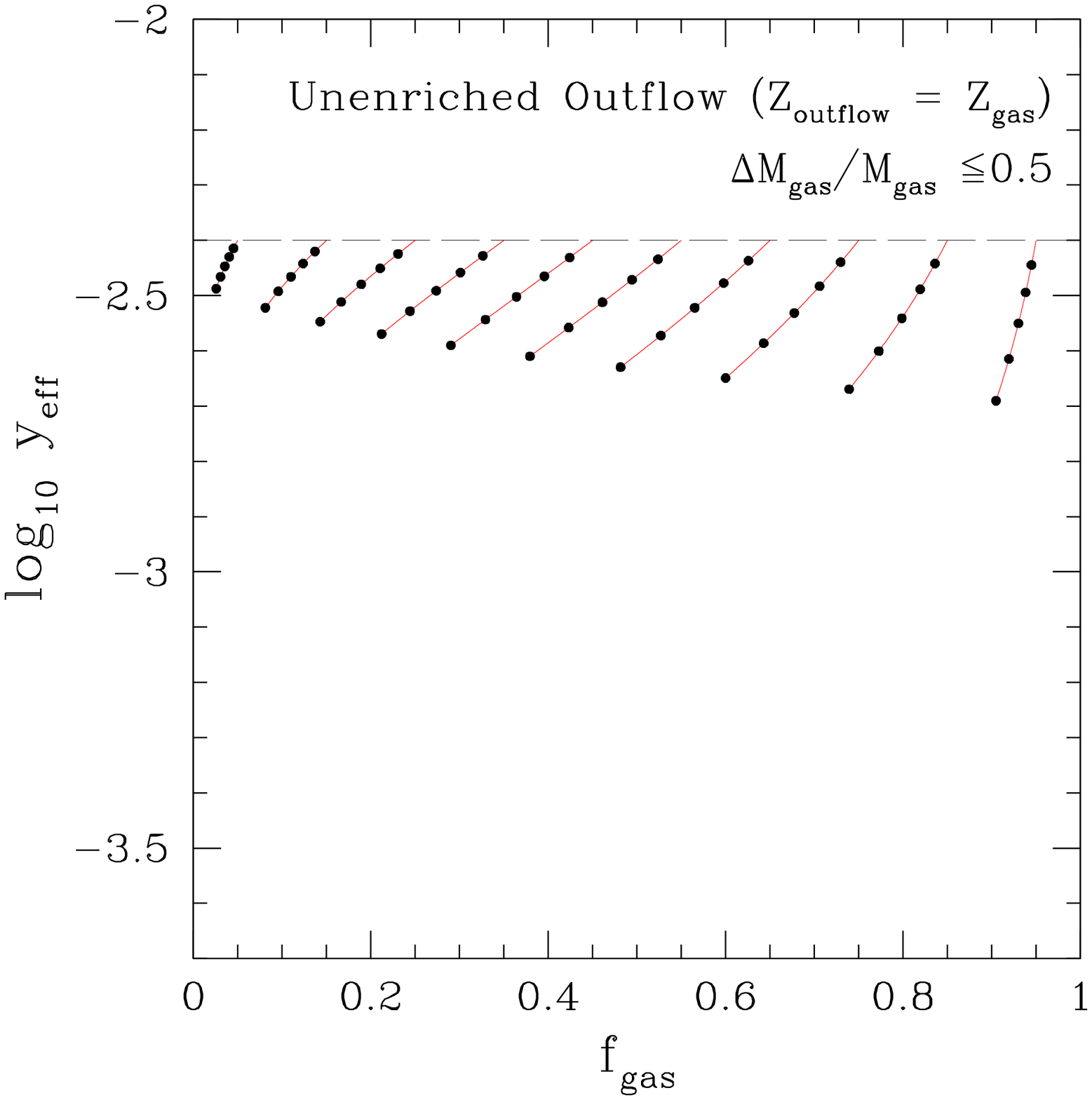}
\includegraphics[width=2.2in]{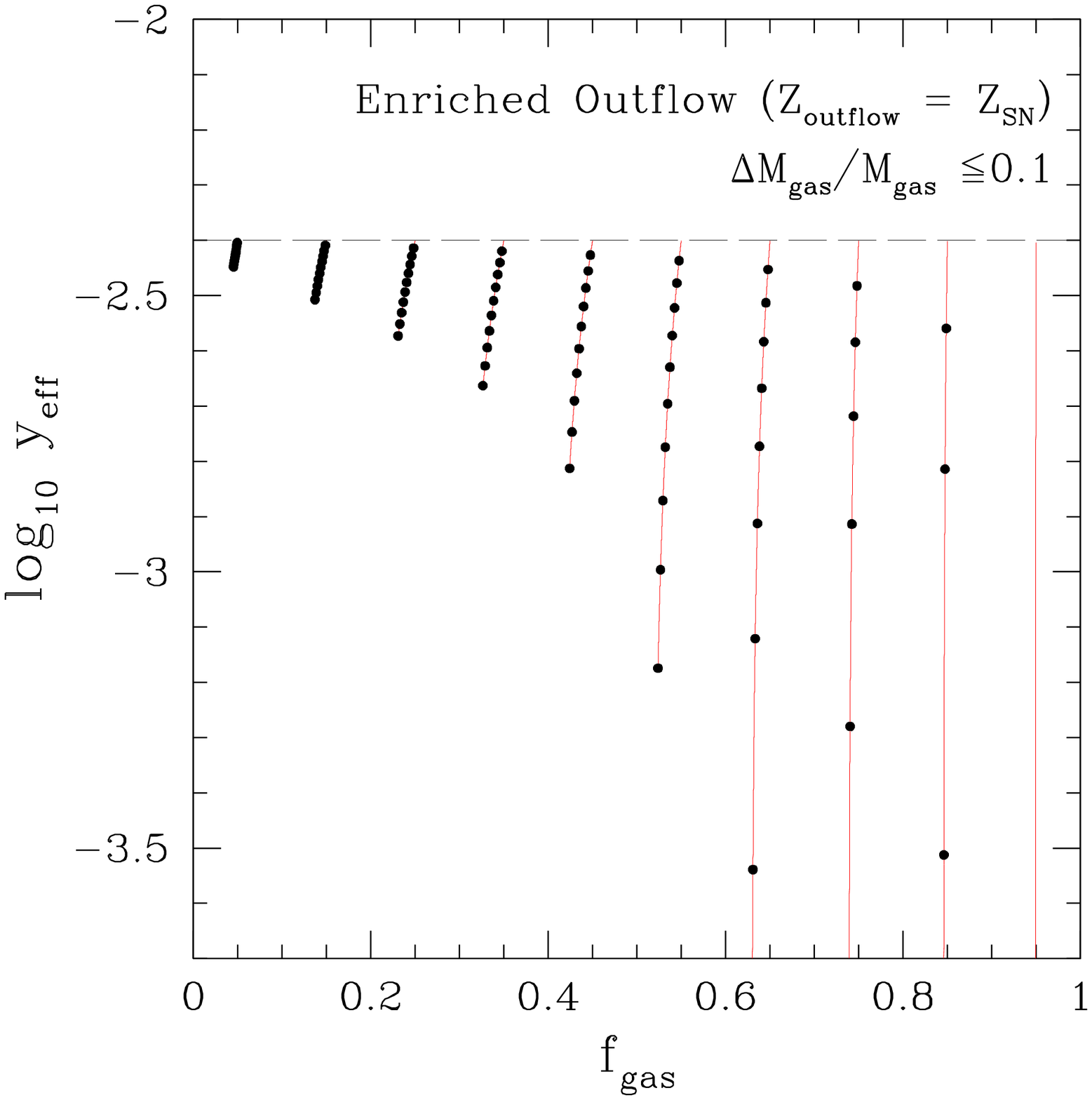}
}
\caption{The evolution of a galaxy's effective yield and gas mass
         fraction in response to unenriched infall ($Z_{infall}=0$;
         left panel), unenriched outflow ($Z_{outflow}=Z_{gas}$;
         center panel), and enriched outflow ($Z_{outflow}=Z_{SN}$),
         for different initial gas mass fractions.  Each galaxy
         initially evolves as a closed-box, with $y_{eff}=y_{true}$,
         indicated with a horizontal dashed line.  Solid lines track
         the evolution of the effective yield and gas mass fraction
         when the initial gas mass is doubled by gas accretion (left
         panel), halved through blast-wave outflow (center panel), or
         reduced by 10\% through enriched outflows (left panel). Dots
         indicate intervals of $\Delta M_{gas}/M_{gas,initial}=0.1$ in
         the left two panels, and 0.01 in the right panel.  Comparing
         to the central panel of Figure~\ref{pilyugindatafig}, neither
         infall nor unenriched outflows can realistically produce the
         very low effective yields seen in dwarf irregular galaxies.
         Simultaneously, the effective yields of more gas-poor,
         massive disks are insensitive to either gas loss or gas
         accretion, and are unlikely to have effective yields that
         deviate strongly from the true nucleosynthetic yield, no
         matter what their evolutionary history.
         \label{yieldfgasfig}}
\end{figure}
\vfill

\begin{figure}[p]
\centerline{
\includegraphics[width=2.2in]{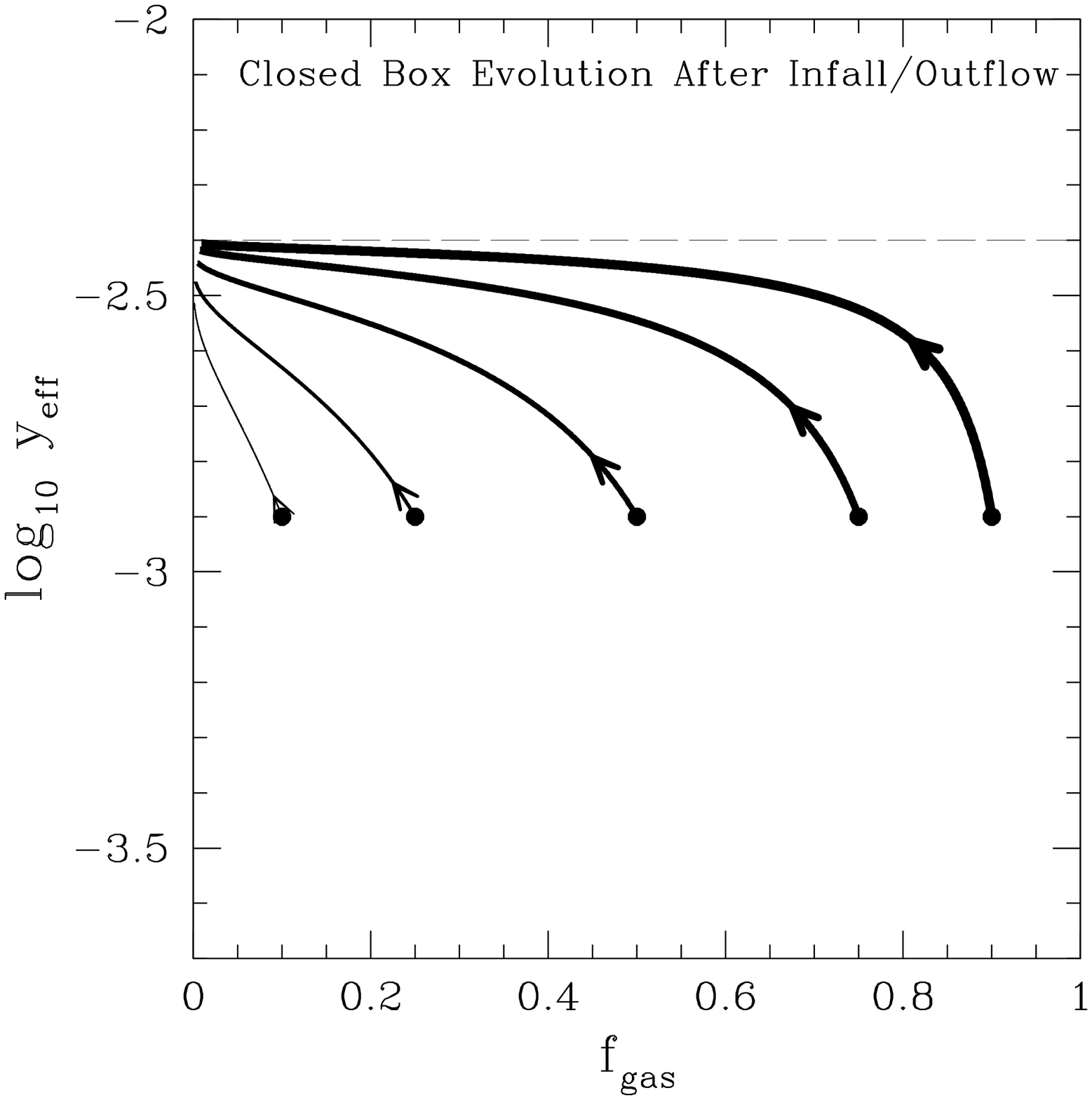}
\includegraphics[width=2.2in]{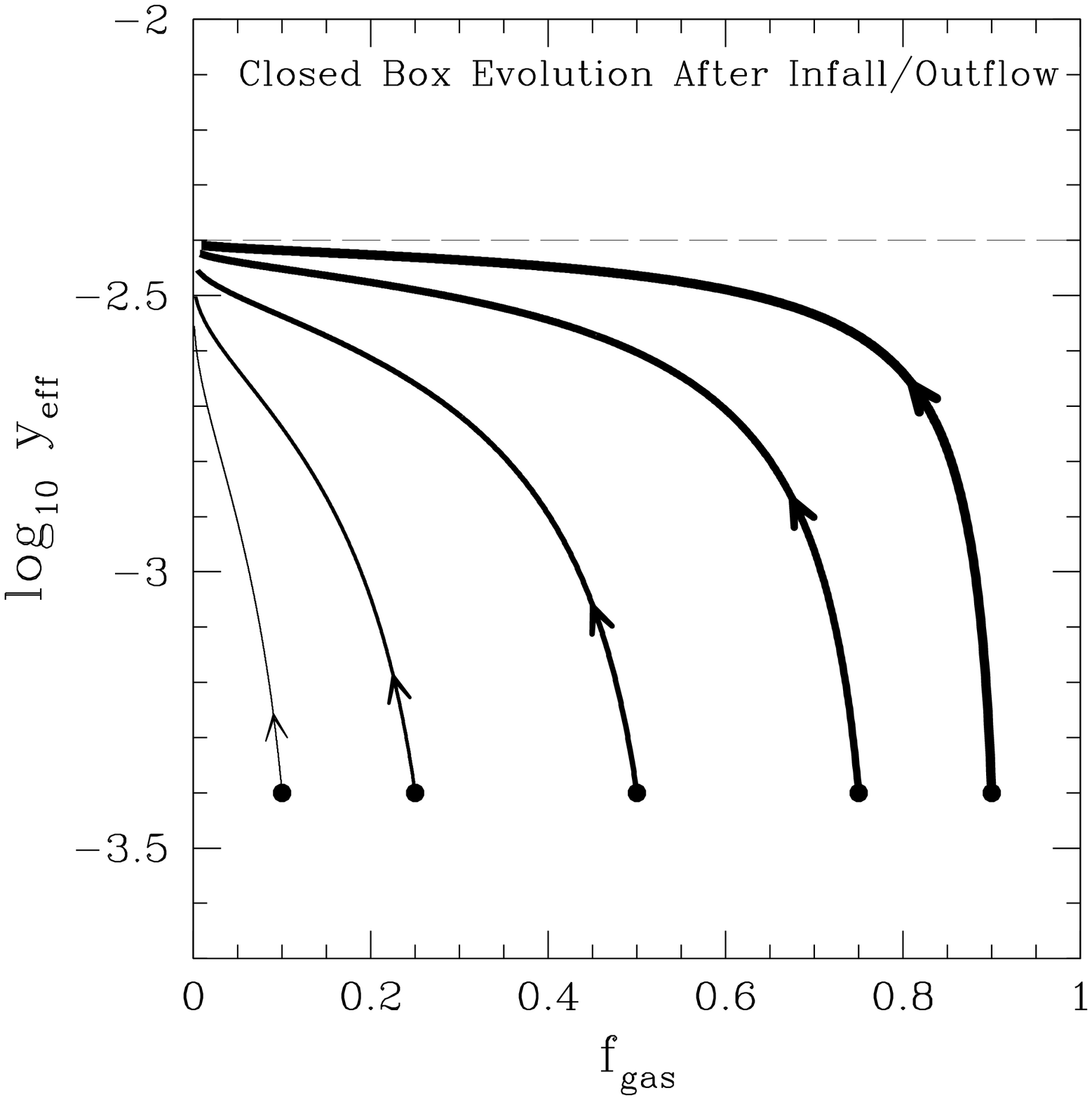}
\includegraphics[width=2.2in]{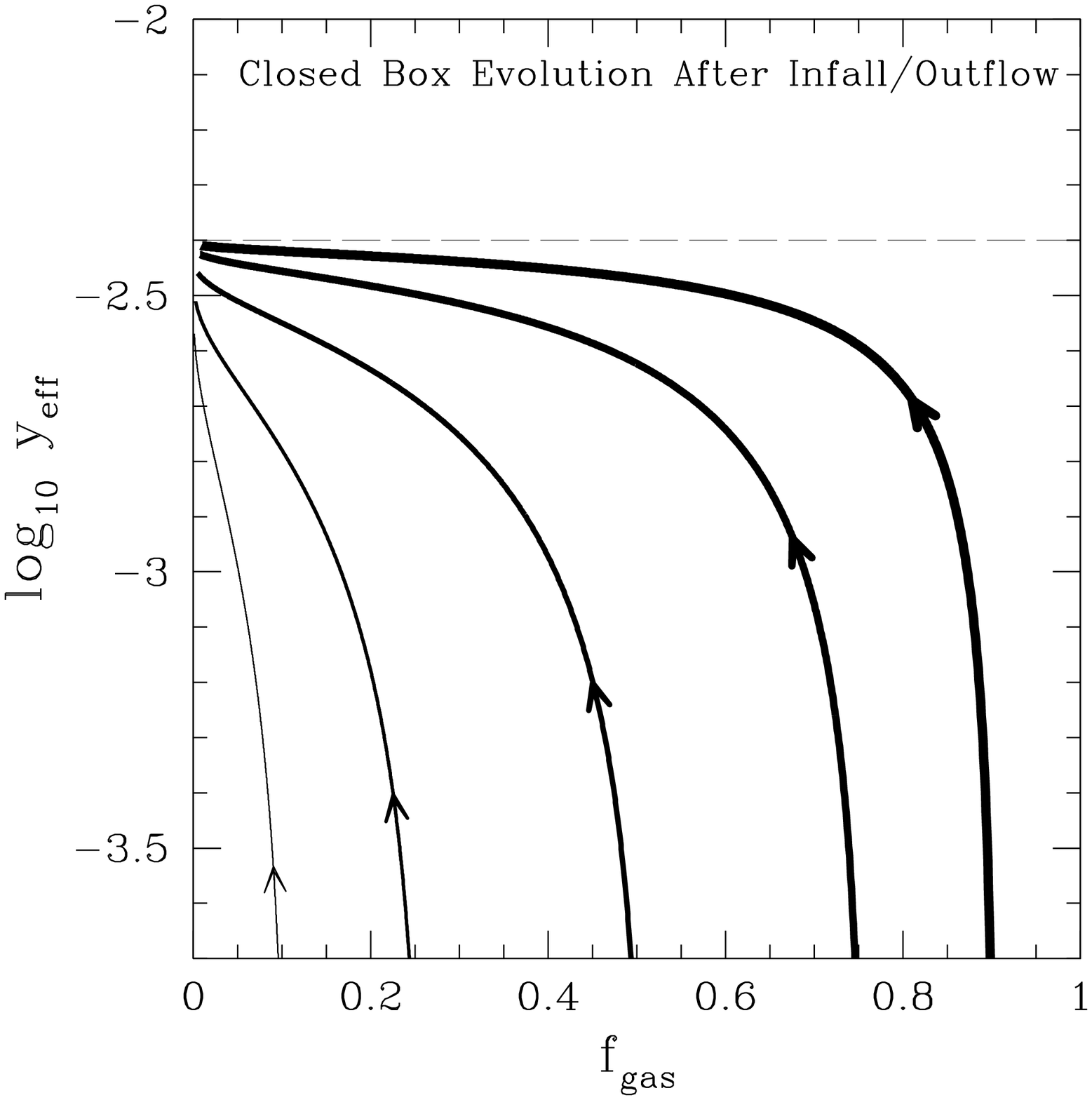}
}
\caption{The evolution of a galaxy's effective yield and gas mass
         fraction during closed-box evolution after a temporary
         reduction in the effective yield, for different initial
         effective yields (${\rm log}_{10} y_{eff,postflow}=-2.9,
         -3.4, -3.9$ in the left, center, and right panels,
         respectively).  In each panel, the different line weights
         indicate different starting gas mass fractions
         ($f_{gas,postflow}=0.1, 0.3, 0.5, 0.75, 0.9$, plotted from
         light to dark (left to right), respectively).  Starting
         values are marked with solid circles, and arrows indicate the
         effective yield after the galaxy consumes 10\% of its gas.
         Galaxies evolve from the right to the left as they convert
         gas into stars, and converge to the true nucleosynthetic
         yield, indicated with a dashed horizontal line.  For low
         starting effective yields (right panel), even small
         reductions in the gas mass fraction cause sharp increases in
         the effective yield.  Thus, low measured effective yields
         require either frequent enriched outflows or inefficient star
         formation.
         \label{evolutionfig}}
\end{figure}
\vfill
\clearpage

\begin{figure}[p]
\centerline{
\includegraphics[width=2.2in]{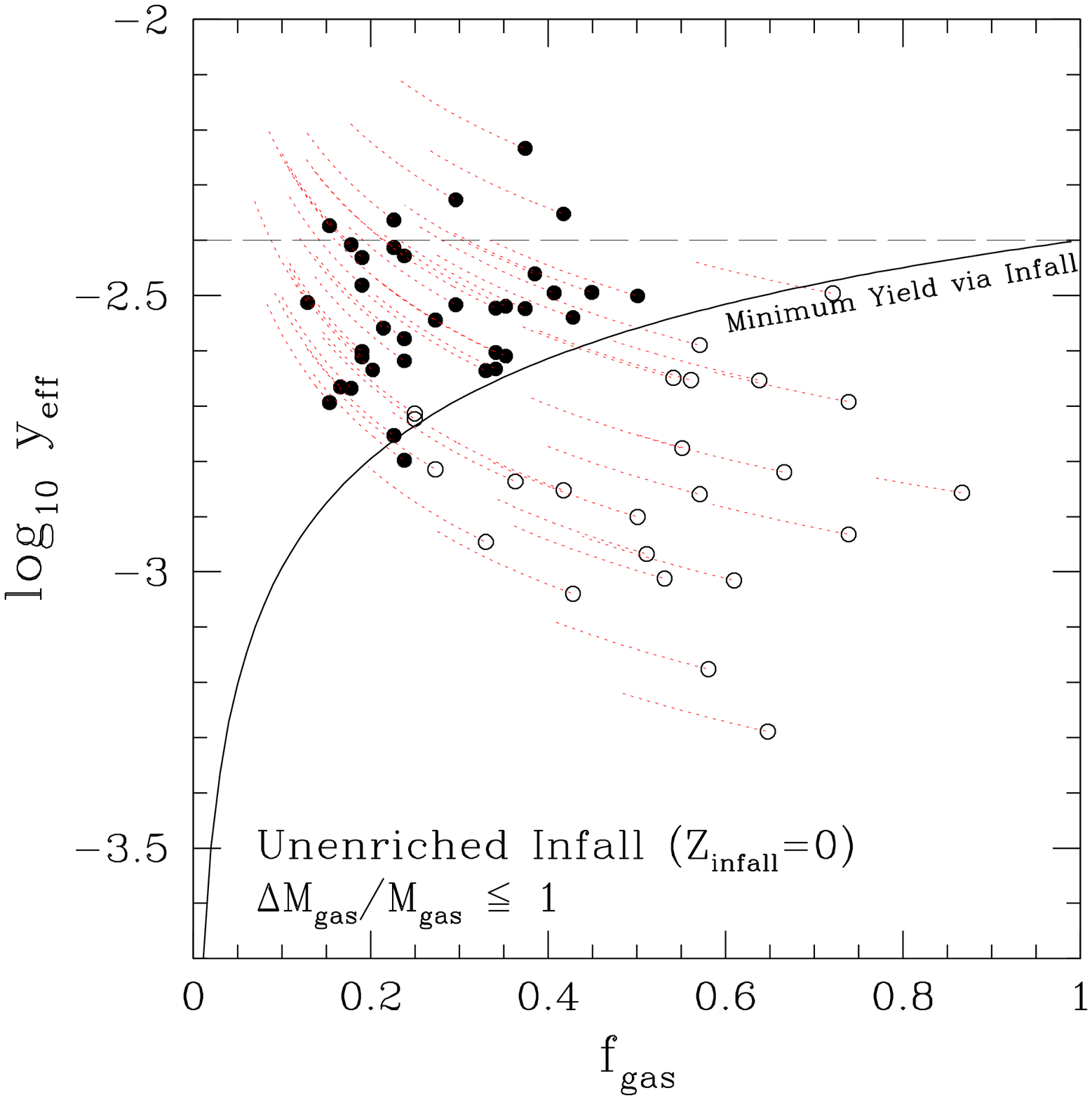}
\includegraphics[width=2.2in]{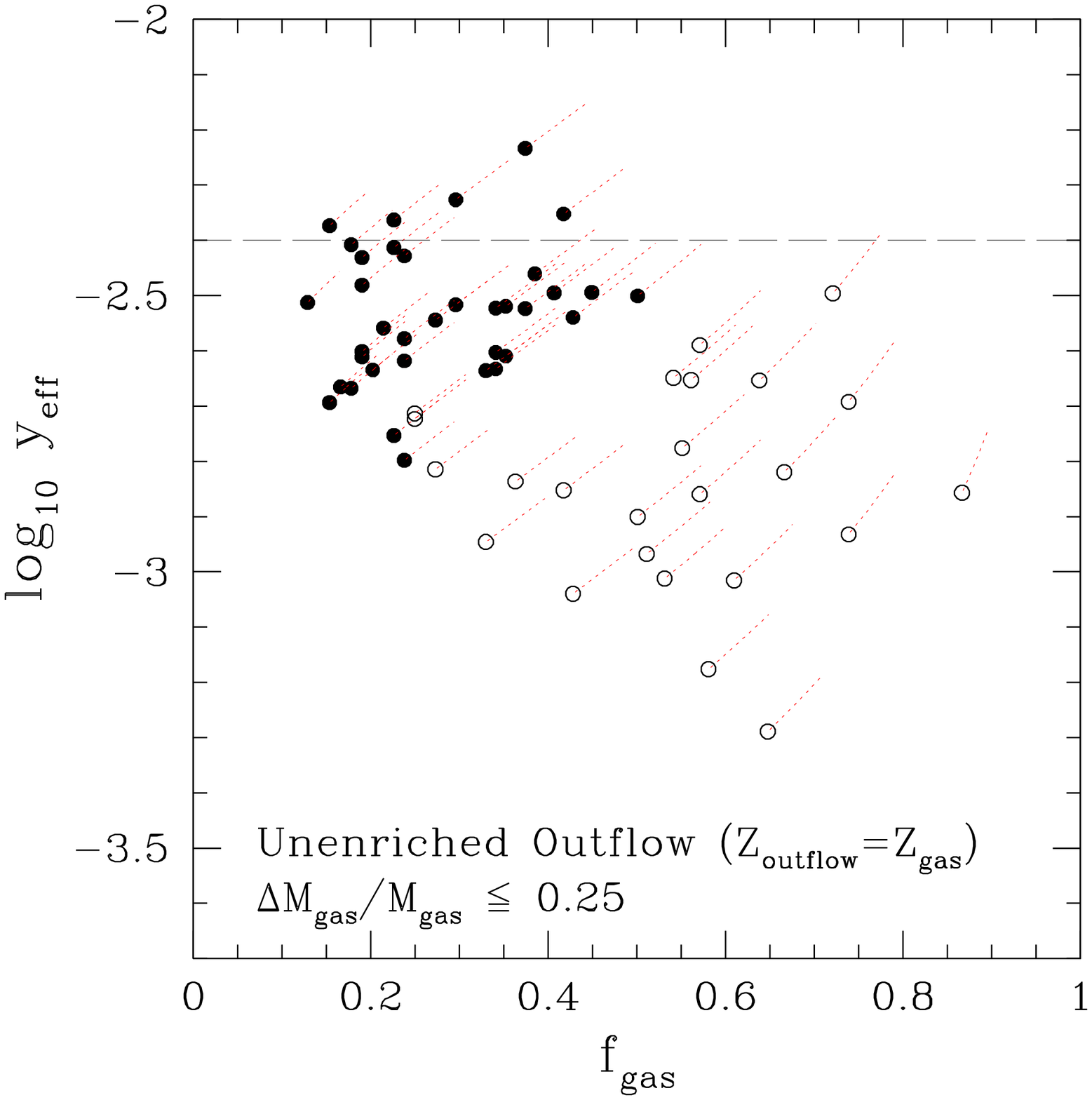}
\includegraphics[width=2.2in]{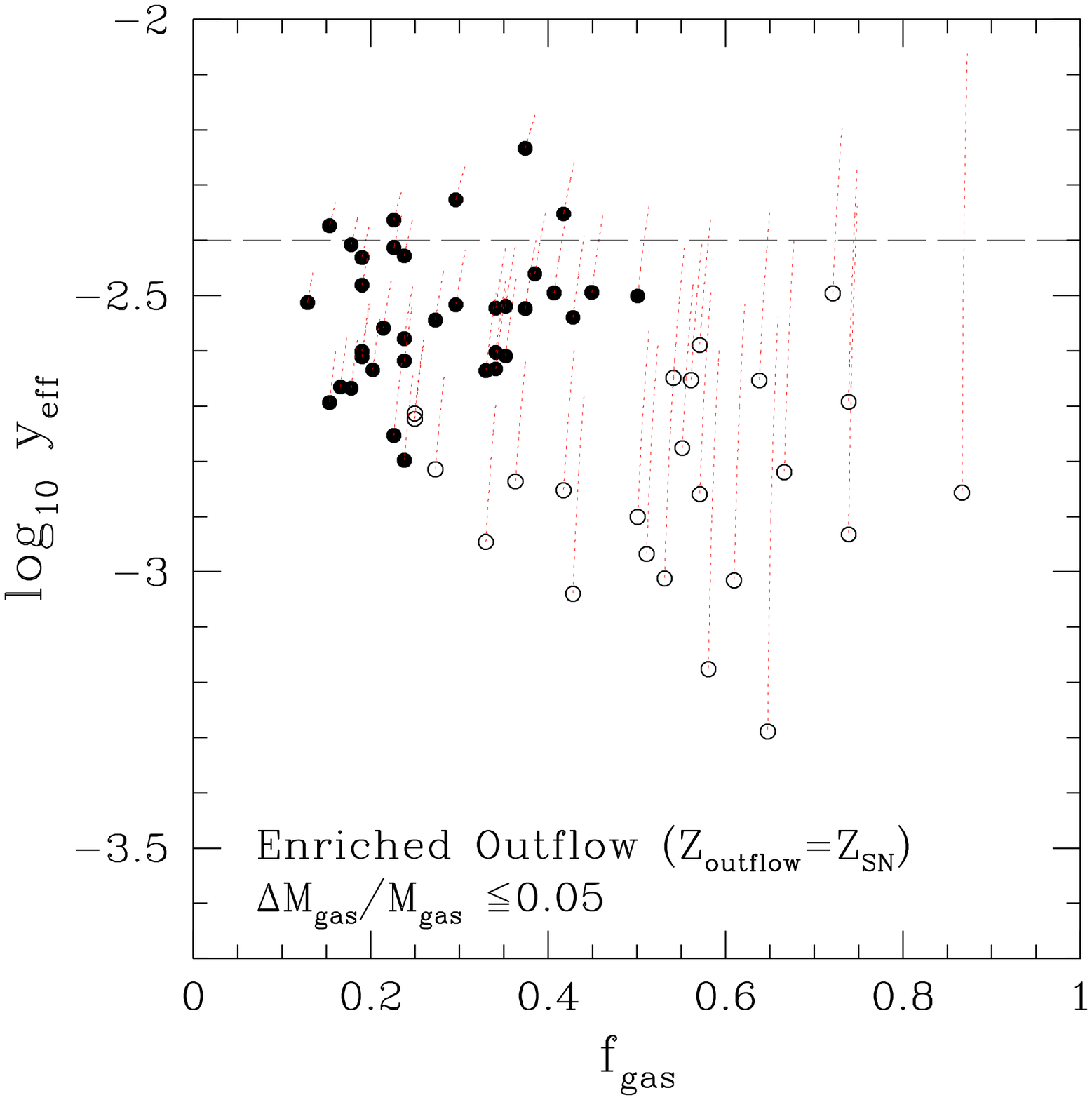}
}
\caption{The effective yield of spirals (solid circles; Sbc-Sd) and
         irregulars (open circles) as a function of the gas mass
         fraction for galaxies from \citet{pilyugin04}.  The dotted
         lines indicate the past effective yield and gas fraction if
         the present state was produced by doubling the gas mass
         through accretion (left panel; $Z_{infall}=0$), by decreasing
         the gas mass by 25\% via unenriched outflows (center panel;
         $Z_{outflow}=Z_{gas}$), or by decreasing the gas mass by 5\%
         via enriched outflows (right panel; $Z_{outflow}=Z_{SN}$).
         Tracks that do not intersect the horizontal dashed line drawn
         at the adopted nucleosynthetic yield $y_{true}$ are not
         likely scenarios for producing the observed data points, or
         require larger gas flows.  Of the three scenarios shown, only
         enriched outflows are capable of producing the lowest
         effective yields with realistic gas flows.  The effective
         yields of gas-poor spirals are insensitive to all outflows,
         though they can be altered significantly by large amounts of
         gas accretion.  In the left hand panel, the solid line shows
         the minimum possible effective yield that infall can produce
         for systems that begin with a given gas mass fraction,
         assuming an initial nucleosynthetic yield of
         $y_{eff,initial}=y_{true}$.
         \label{pilyuginfig}}
\end{figure}
\vfill

\begin{figure}[p]
\centerline{
\includegraphics[width=5in]{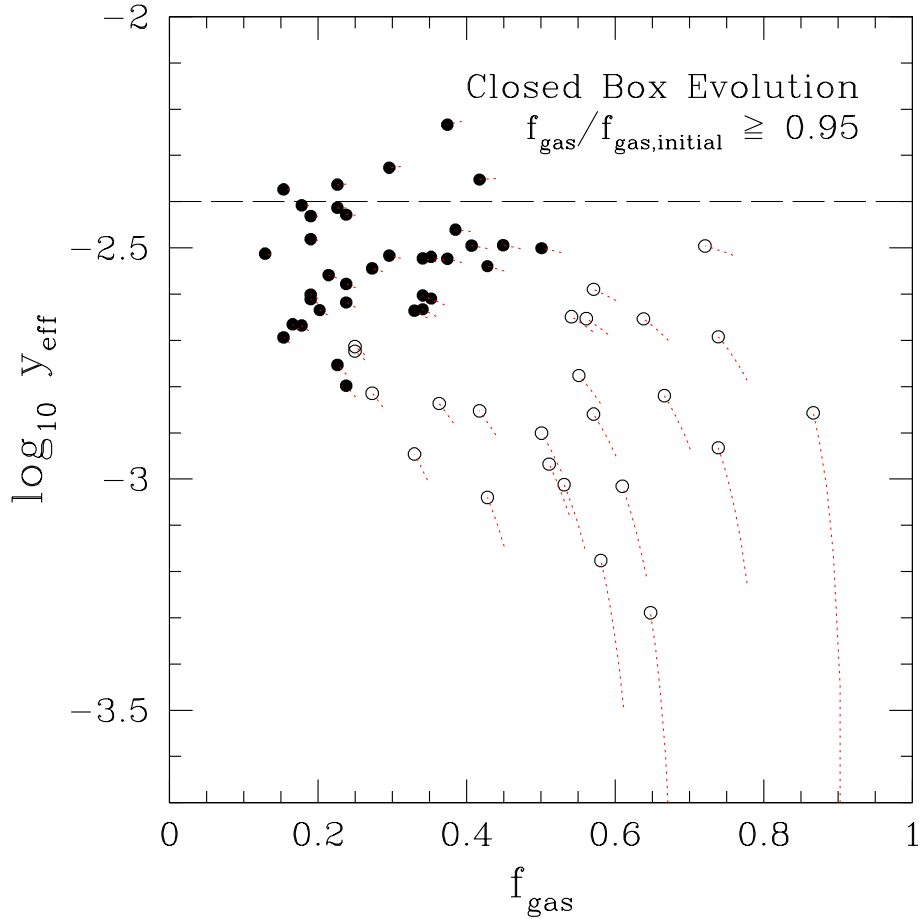}
}
\caption{The effective yield of spirals (solid circles; Sbc-Sd) and
         irregulars (open circles) as a function of the global gas
         mass fraction for galaxies from \citet{pilyugin04}.  The
         dotted lines indicate the past effective yield and gas
         fraction if the present values were produced by closed-box
         evolution that takes place after outflow or infall has
         stopped.  This subsequent evolution reduces the gas mass
         fraction and pushes the effective yield back to the assumed
         nucleosynthetic yield, as shown in
         Figure~\ref{evolutionfig}. The length of the line assumes
         that the galaxy has converted only 5\% its initial gas into
         stars to reach the present value.  Even assuming this very
         modest evolution, the galaxies with the lowest effective
         yields and/or largest gas mass fractions would have had
         substantially lower effective yields in the past.
         \label{pilyuginevolutionfig}}
\end{figure}
\vfill
\clearpage

\begin{figure}[p]
\centerline{
\includegraphics[width=3.2in]{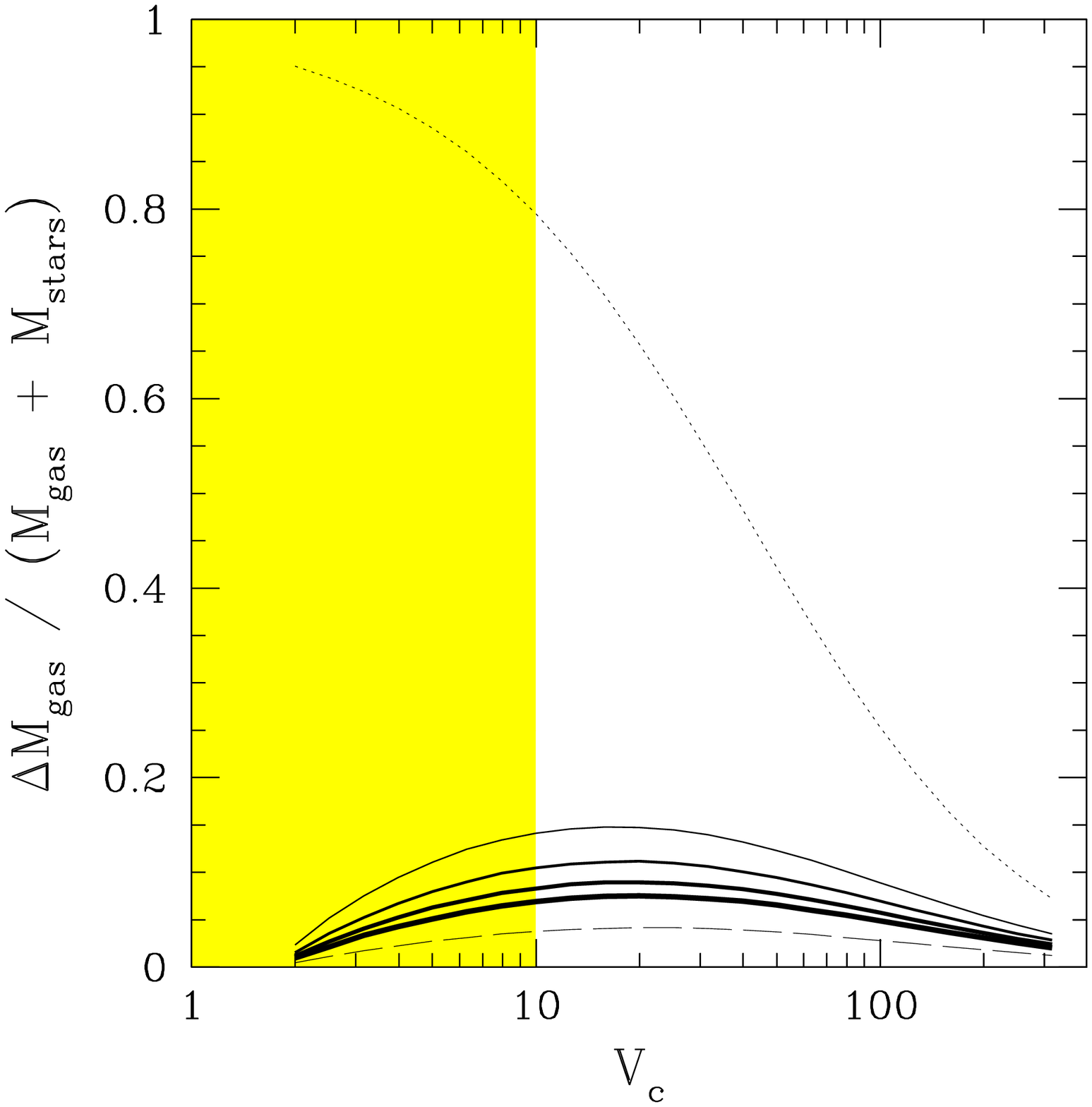}
\includegraphics[width=3.2in]{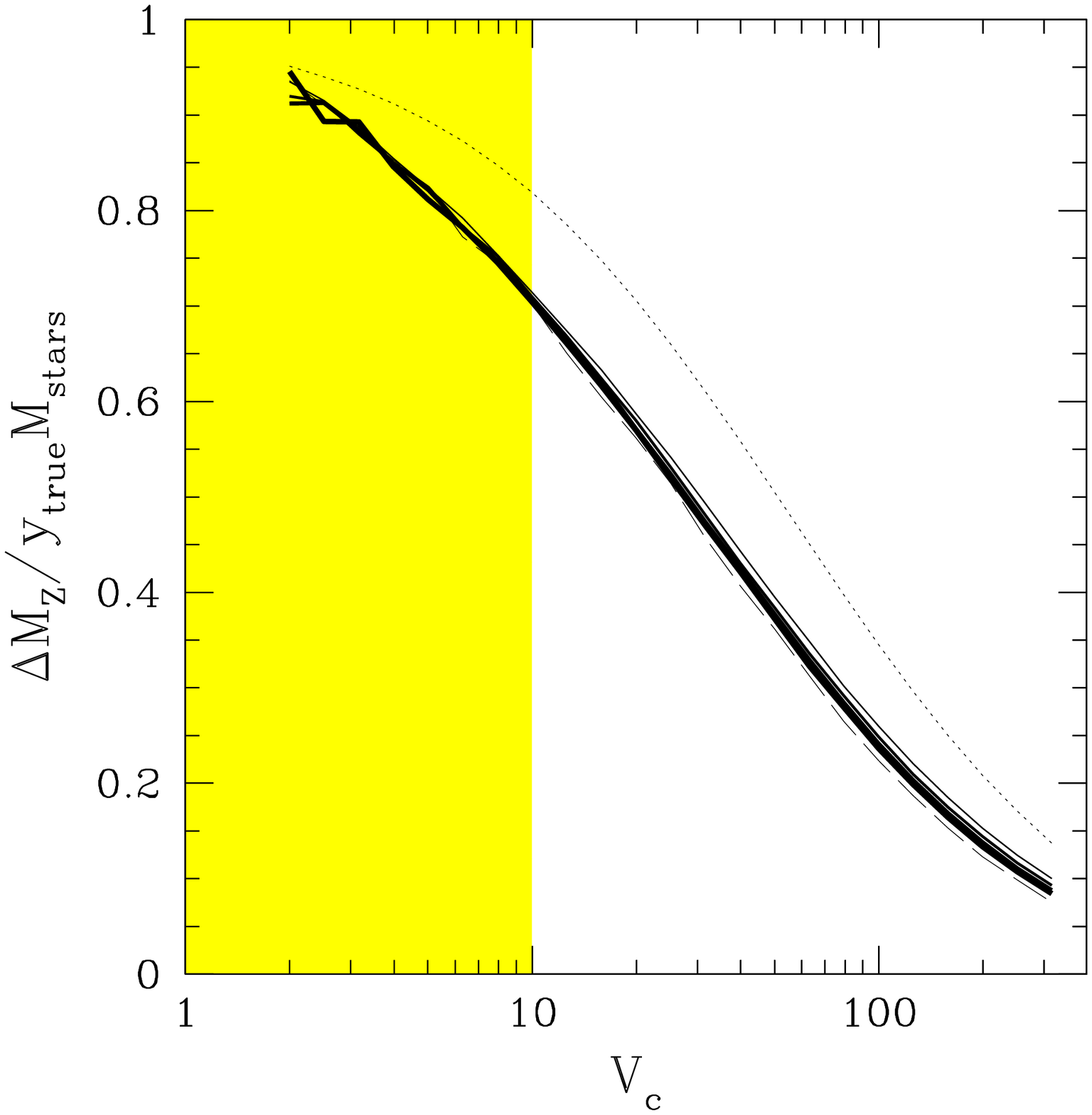}
}
\caption{The upper limit to the mass fraction of baryons lost (left panel)
  and metals lost (right panel) as a function of galaxy rotation
  speed.  The lines indicate different fractions of 
  entrained ISM, ranging from 0\% (pure SN ejecta; lower
  dashed line) to 100\% (pure blast wave; upper dotted line), with
  intermediate values of $\epsilon=0.5-0.8$ in steps of 0.1, plotted
  as solid lines from dark to light.  The models assume that no infall
  has taken place, and that the galaxies experienced a single outflow
  event following a period of closed box evolution.  These assumptions
  guarantee that the mass loss fractions are the maximum that could be
  obtained for an arbitrary star formation history involving the same
  initial and final gaseous and stellar masses.  The shaded area
  indicates the region where models are unconstrained by current data.
  \label{modelfig}}
\end{figure}
\vfill

\begin{figure}[p]
\centerline{
\includegraphics[width=2.2in]{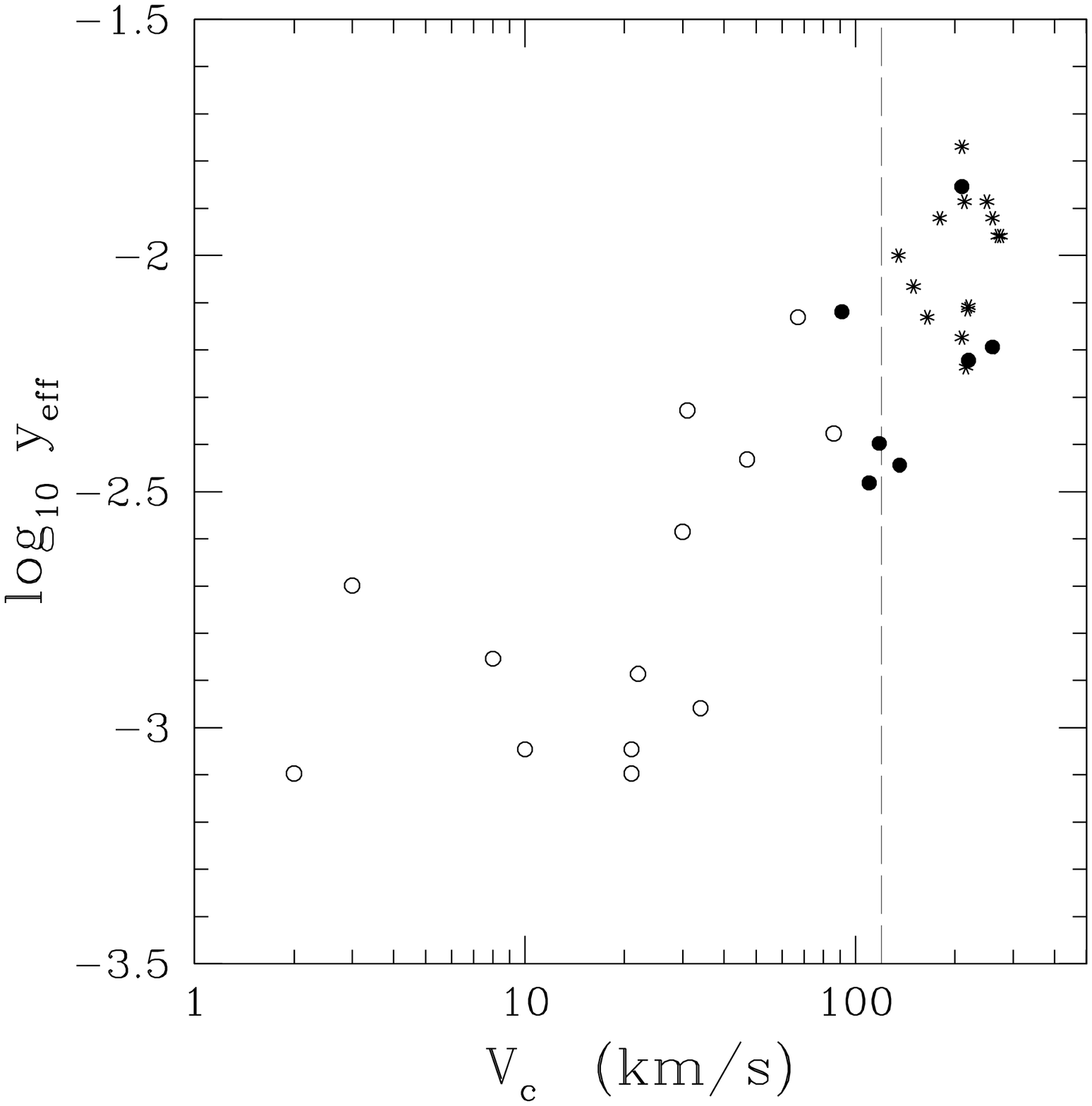}
\includegraphics[width=2.2in]{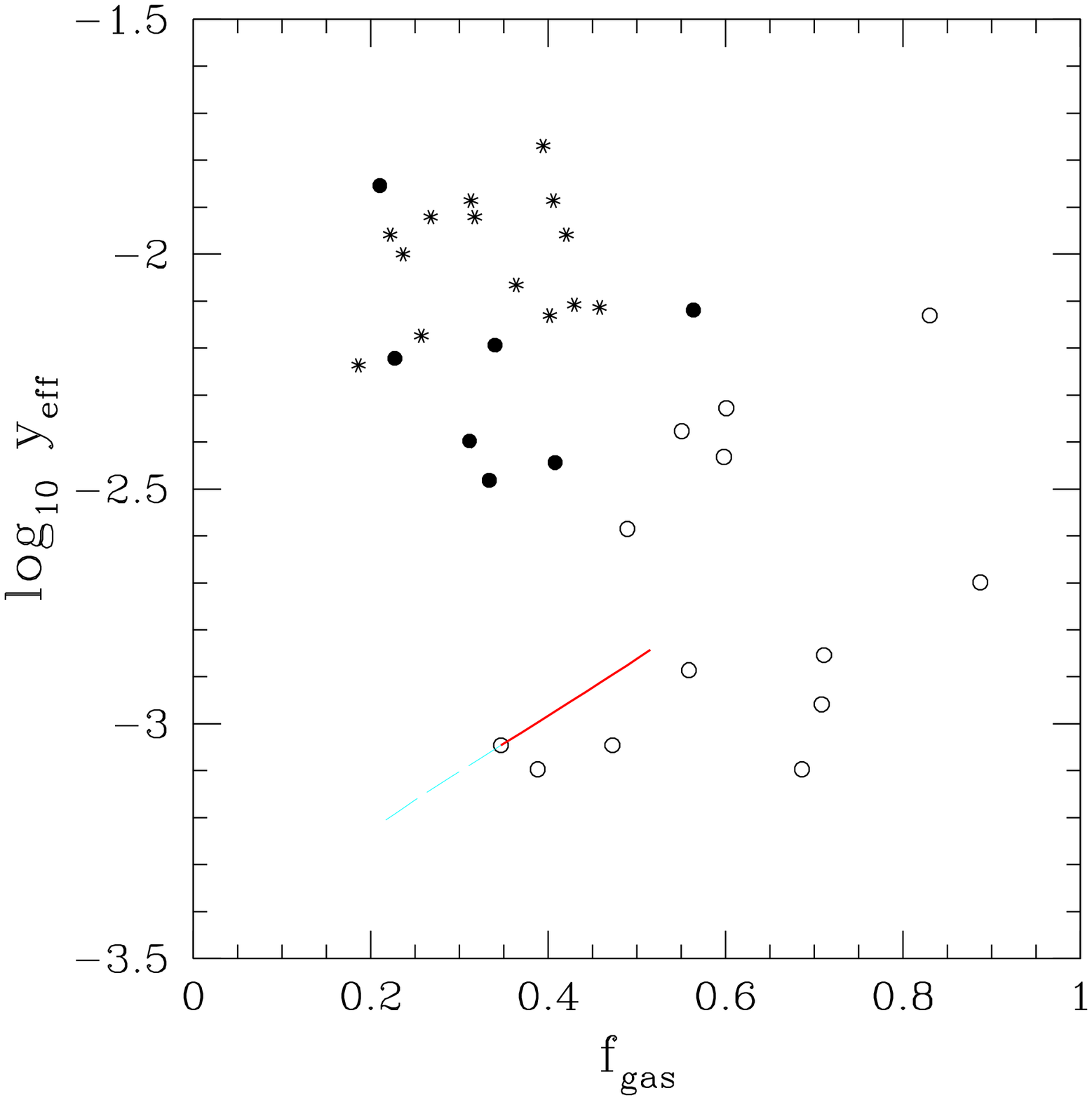}
\includegraphics[width=2.2in]{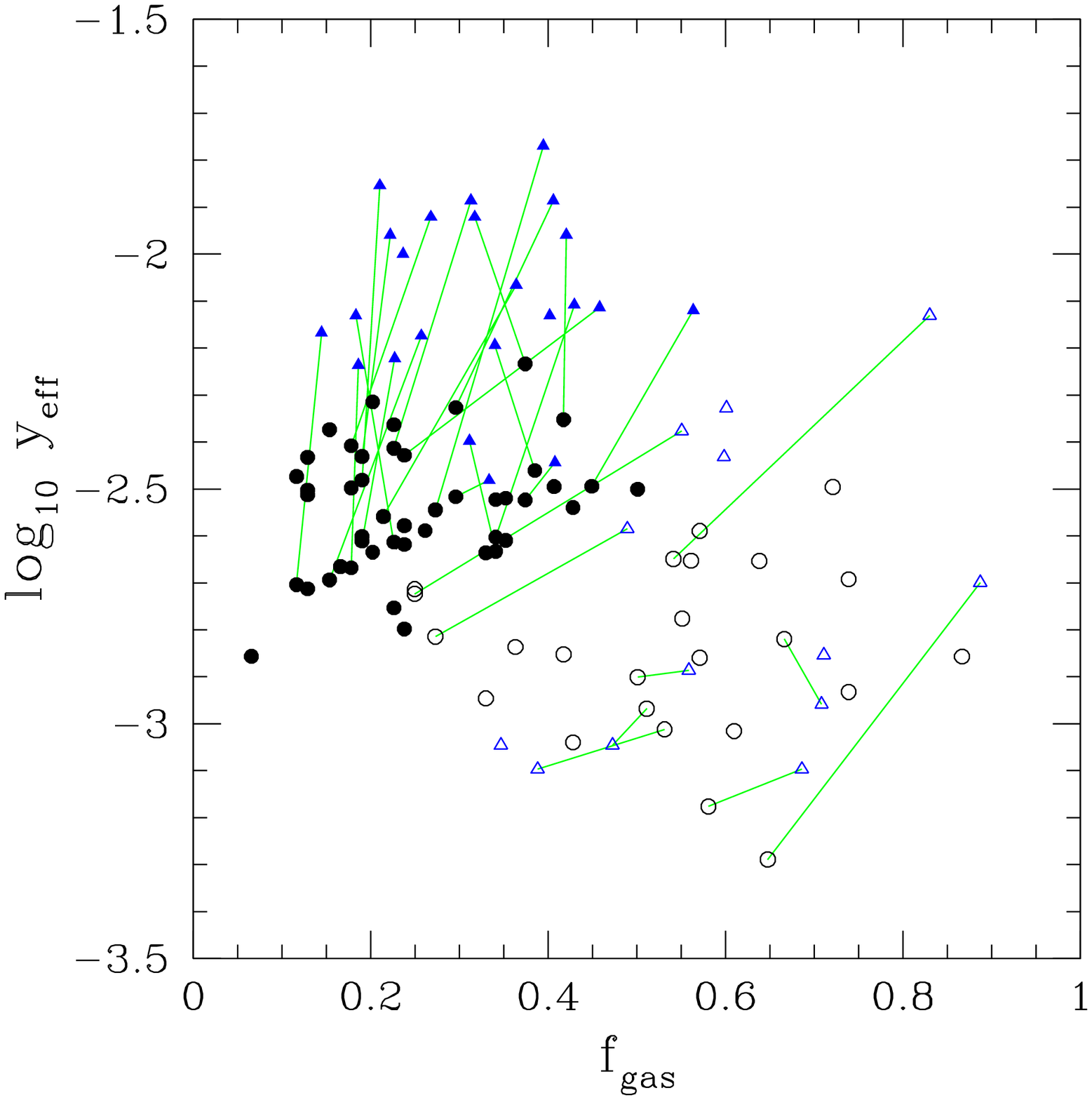}
}
\caption{Effective yield at the disk half-light radius as a function
         of galaxy rotation speed (left panel) and gas mass fraction
         (center panel), for the data from Table~4 of
         \citet{garnett02}.  Mid-type spirals (Sbc-Sc) are plotted as
         stars, late-type spirals (Scd-Sd) as solid circles, and
         irregulars as open circles. Only types Sbc and later are
         included.  All lines in the plots are equivalent to those
         shown for the \citet{pilyugin04} data in the left two plots
         of Figure~\ref{pilyugindatafig}.  The right panel compares
         the data from \citet{garnett02} (triangles) to
         \citet{pilyugin04} (circles), with lines connecting the
         different measurements reported for identical galaxies.  The
         assumptions used in \citet{garnett02} result in larger
         effective yields and gas mass fractions.  See the discussion
         in Appendix~\ref{garnettsec} for the origin of these
         differences.
	 \label{garnettdatafig}}
\end{figure}
\vfill
\clearpage

\begin{figure}[p]
\centerline{
\includegraphics[width=3.2in]{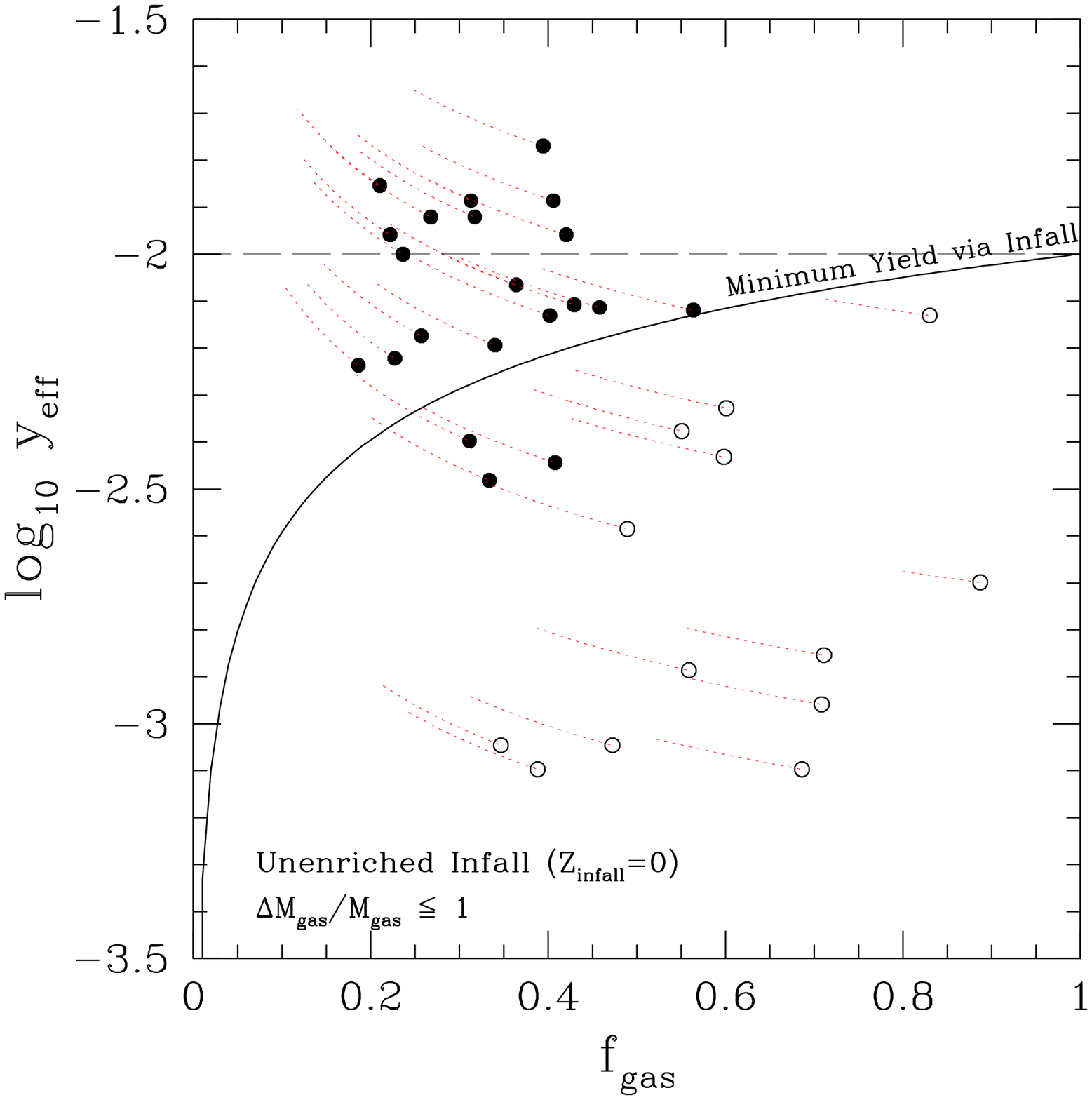}
\includegraphics[width=3.2in]{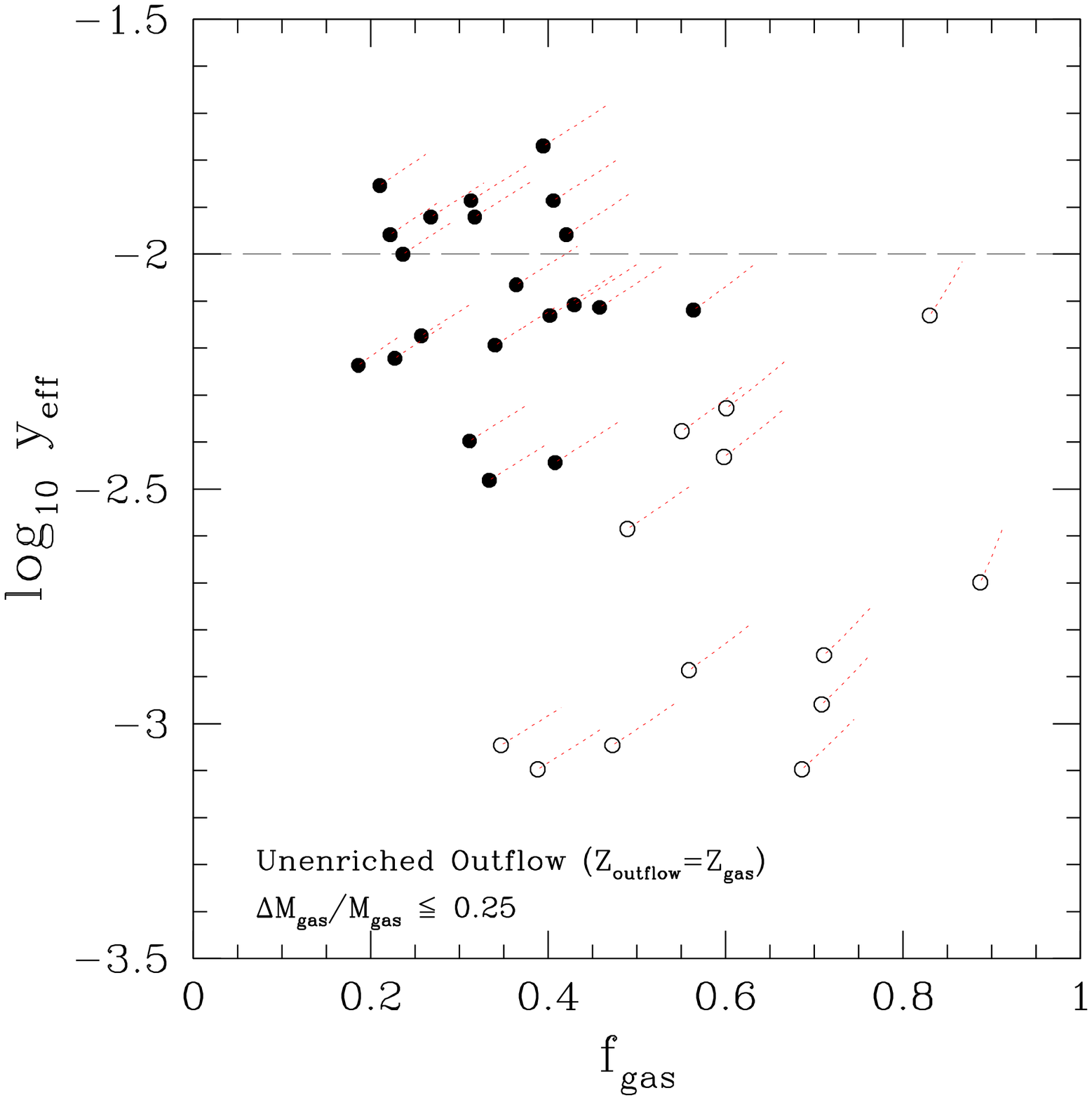}
}
\centerline{
\includegraphics[width=3.2in]{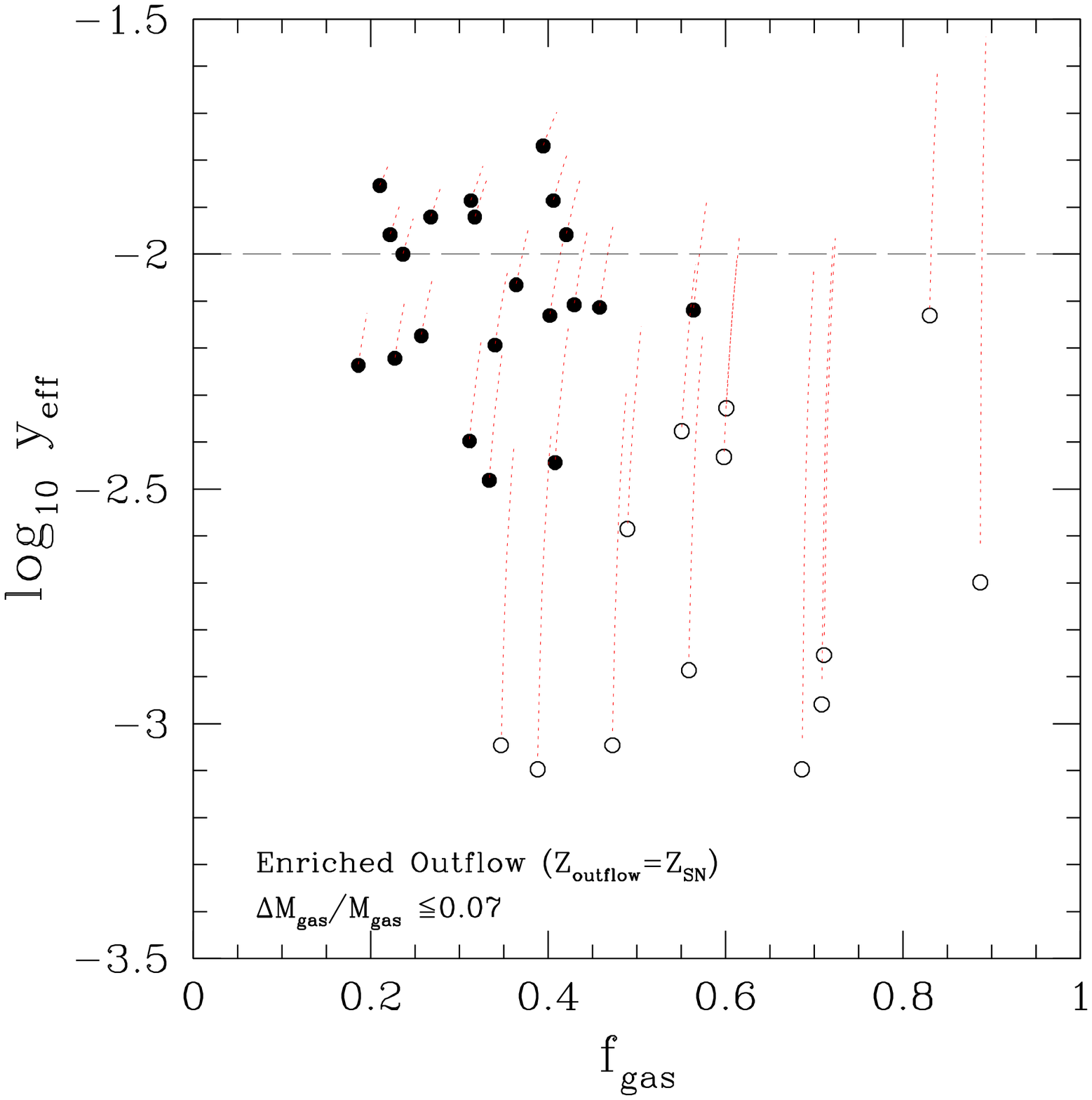}
\includegraphics[width=3.2in]{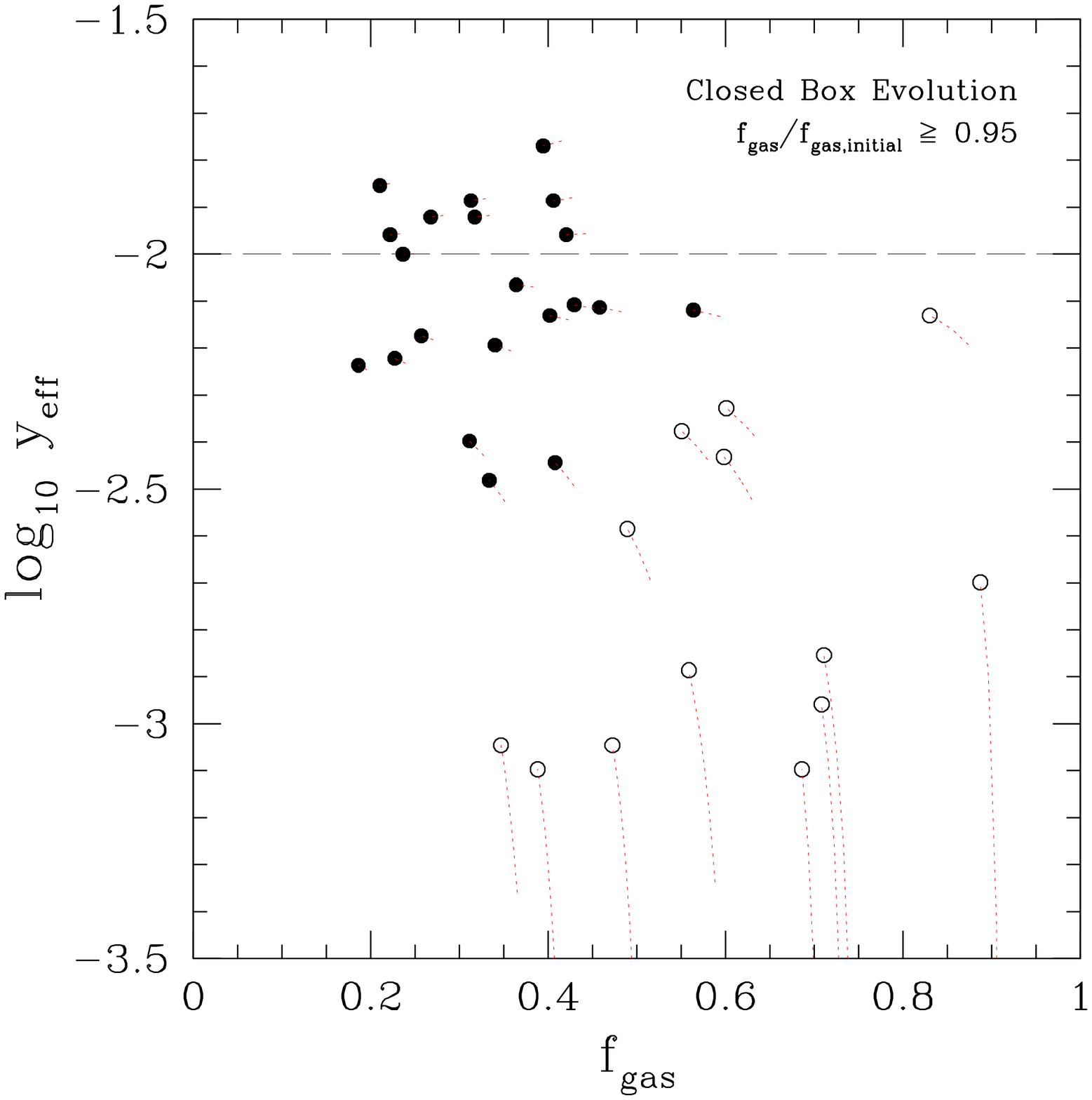}
}
\caption{The effective yield of spirals (solid circles) and irregulars
         (open circles) as a function of the gas mass fraction for
         galaxies from \citet{garnett02}.  The dotted tracks are
         equivalent to those in
         Figures~\ref{pilyuginfig}~\&~\ref{pilyuginevolutionfig} for
         unenriched infall (upper left), unenriched outflow (upper
         right), enriched outflow (lower left), and subsequent
         closed-box evolution (lower right).
         \label{garnettfig}}
\end{figure}
\vfill
\clearpage

\fi

\end{document}